\documentclass[jws,biber]{template/journal}
\usepackage[utf8]{inputenc}

\title{Characterizing the country-wide adoption and evolution of the Jodel messaging app in Saudi Arabia}

\author[1]{Jens Helge Reelfs}
\author[1]{Oliver Hohlfeld}
\author[2]{Markus Strohmaier}
\author[3]{Niklas Henckell}
\affil[1]{Brandenburg University of Technology, reelfs@b-tu.de, hohlfeld@b-tu.de}
\affil[3]{Business School, University of Mannheim, markus.strohmaier@uni-mannheim.de}
\affil[4]{The Jodel Venture GmbH, niklas@jodel.com}

\articledatabox{%
ISSN 2332-4031; DOI 10.34962/jws-107\\
\copyright\ 2022 J. H. Reelfs, O. Hohlfeld, M. Strohmaier, N. Henckell}

\issuecopyrightyear{2022}
\issuevolumenumber{8}
\issuenumber{x}
\issuevolumeyear{2022}
\issuestartpage{1}

\usepackage{url}  %
\usepackage{graphicx}  %
\usepackage{booktabs} %

\usepackage{tikz}
\usepackage{pgfplots}
\pgfplotsset{compat=1.13}
\usepgfplotslibrary{colormaps}
\usetikzlibrary{pgfplots.colormaps}
\usetikzlibrary{pgfplots.dateplot}
\usetikzlibrary{patterns}
\usepackage{graphicx}
\usepackage{subcaption}
\usepackage{color, colortbl}
\usepackage{arydshln}
\usepackage{float}
\usepackage{arabtex}
\usepackage{utf8}
\usepackage{units}
\usepackage{enumitem}
\usepackage{balance}
\usepackage{microtype} %
\usepackage[export]{adjustbox}

\setcode{utf8}

\definecolor{Gray}{gray}{0.9}

\newcommand{\afblock}[1]{\vspace*{.2em}\noindent{\textbf{#1 }}}
\newcommand{\takeaway}[1]{\vspace*{.2em}\noindent{\textbf{Findings.}} \textit{#1}\vspace*{.2em}}

\newcommand\encircle[1]{%
	\tikz[
		baseline={([yshift=-8pt]current bounding box.north)}
	]
		\node (X) [draw, shape=circle, inner sep=0, fill=black, text=white,scale=0.7] {\strut #1};
	\hspace{-4pt}
}

\newcommand{\ie}{i.e.,\,}
\newcommand{\Ie}{I.e.,\,}
\newcommand{\eg}{e.g.,\,}

\newcommand{\etal}{et~al\@ifnextchar.{}{.\@}}
\newcommand{\etc}{etc\@ifnextchar.{}{.\@}}

\newcommand{\wrt}{w.r.t.\@}
\newcommand{\sref}[1]{\S~\ref{#1}}

\begin{document}
\section*{Abstract}

Social media is subject to constant growth and evolution, yet little is known about their early phases of adoption.
To shed light on this aspect, this paper empirically characterizes the initial and country-wide adoption of a new type of social media in Saudi Arabia that happened in 2017.
Unlike established social media, the studied network Jodel is anonymous and location-based to form hundreds of independent communities country-wide whose adoption pattern we compare.
We take a detailed and full view from the operators perspective on the temporal and geographical dimension on the evolution of these different communities---from their very first the first months of establishment to saturation.
This way, we make the early adoption of a new type of social media visible, a process that is often invisible due to the lack of data covering the first days of a new network.


\newcommand{\numberNumCharactersPerJodel}[0]{$250$}
\newcommand{\numberMarchPeakRegistrations}[0]{$28\text{k}$} %
\newcommand{\numberMarchBeginningPosts}[0]{$193\text{k}$} %
\newcommand{\numberMarchPeakPosts}[0]{$461\text{k}$} %
\newcommand{\numberMarchPeakReplies}[0]{$2.5\text{M}$} %
\newcommand{\numberMarchPeakRepliesToPosts}[0]{$5.5$} %
\newcommand{\numberMarchPeakUpvotes}[0]{$3.7\text{M}$} %
\newcommand{\numberMarchPeakDownvotes}[0]{$1.3\text{M}$} %
\newcommand{\numberMarchPeakVoteRatio}[0]{$74\%$} %
\vspace{-1.5em}
\section{Introduction}
Jodel or Whisper are examples of a new class of emerging anonymous, location-based web applications that \emph{i)} enable users to post anonymously, without displaying user-related information and \emph{ii)}~display content only in the proximity of the user's location.
In case of Jodel---a phone based web app for anonymous location-based messaging---posts are only displayed to other users within close (up to 20\,km) geographic proximity.
This location-based nature of \emph{only} displaying nearby content and the inability to communicate countrywide results in the formation of hundreds of \emph{independent} communities throughout a country.
Consequently, adoption patterns can vary between different cities or rural environments, opening questions on how adoption spreads and the app usage diffuses through a complete country.

Despite the gaining popularity of this novel type of apps, little is known about how they are adopted, or what factors drive adoption and if adoption (success) might be controllable.
In particular, the early phases of app adoption are understood poorly even though understanding the social mechanisms behind such diffusion processes are crucial for the design and roll-out of such platforms.
This lack of information on early phases of adoption is rooted in a lack of empirical data covering the early phases of a new platform.

This paper presents the first empirical characterization of the nation-scale adoption of an anonymous, location-based messaging app in the KSA, covering the entire adoption phase beginning with the first registered user in March 2015 up until saturation in 2017.
Note that as of today, the platform is still in use within the KSA.
Given that research on these phenomena relies on the cooperation with its operators, this kind of country-wide studies have not been broadly available to our community so far.
Our observation period includes the time from the first registered user in March~2015 to the country-wide establishment in August~2017.
We focus, however, on the time from the first significant app interactions within the KSA in Aug~2016 until the beginning of Aug~2017.
The data spans about 1~billion events including data from about 1~million users and 500~million posts.
While adoption patterns can differ for different apps and countries, in the absence of data on this matter, we present a first characterization of nation scale adoption processes to enable and model user behavior. %

In March 2017 the application experienced a sudden and drastic influx of new users where the usage increased from hardly any to country-wide adoption.
From a network perspective, this sudden adoption of a new application represents a change in network traffic in which a new application suddenly appears at a country-wide scale.
From an operator perspective, this sudden adoption can be initially looking like malicious use (\eg{} to spam the application).

\afblock{Research objectives and questions.}
The aim of this paper is to \emph{empirically characterize} and \emph{model} the early adoption phase of new social media \wrt{} to user adoption behavior.
Jodel is a well-suited network to study this question given the fact that its location-based nature---in which no country-wide communication is possible---enables us to compare the behavior of hundreds of \emph{independent} communities country-wide.
The establishment of these communities raises the general question if and how they adopt and evolve over time and with size.
While focussing on various aspects from a community-view and then switch to a user-oriented perspective, our study is driven by three essential questions.

I)~We shed light on the temporal and geographic development of communities on a national level~(\sref{sec:adoption}): how fast did the adoption in the KSA occur and do adoption pattern differ by community?
II)~We provide insights into overall community activity.
That is, in \sref{sec:community} we discuss measures \wrt{} main app characteristic: content feeds via
\emph{a1)} content creation \& content appreciation, 
\emph{a2)} platform behavior \wrt{} response times, discussion homogeneity \& vote-consensus, 
and finally in \emph{b)}, we elaborate on new user influx and user interactions.
III)~We aim to understand if the different Jodel communities differ \wrt{} user behavior~(\sref{sec:user})---in particular 
\emph{a)} How diverse is the set of participating users? 
Moreover, how can we measure implications on users of key app design decision being \emph{b1)} anonymous \& \emph{b2)} location-based.
Lastly, \emph{c)} we ask how many active users do the communities attract, how long users keep using the platform, and what makes users leaving the platform?

%
%
%
	%
	%
	%

Our study is driven by the comparison of a plethora of independent communities country-wide by focusing on \emph{i)} their temporal and geographic adoption, \emph{ii)} a community driven analysis to understand interlinked interactions, and \emph{iii)} a user's perspective of participation and platform experience.
By comparing these communities \wrt{} different properties, we provide models where applicable and thus largely identify similarities, scaling effects and differences at times.
This way, we can empirically characterize and model the adoption and usage of a new type of social media on a country wide scale.
This presents the first comprehensive analysis of a new social messaging app that combines location-based content with the ability to post anonymously.
It has large user-bases in Europe, the US, and as we show, a sudden adoption in Arabic speaking countries.

\afblock{Contributions and findings.}
\begin{itemize}[noitemsep,topsep=5pt,leftmargin=9pt]
	
	\item 
	Based on data provided by the network operator, we trace the birth of its adoption within the KSA.
	Its adoption began in Riyadh (capital) and happened in 3 phases, most notably a phase of sudden adoption---in all communities simultaneously---supported by social media influencers advertising Jodel.
	Surprisingly, the different communities show the same qualitative adoption pattern nationwide.
    
    \item 
	We characterize interaction distributions across the communites influencing the major app design decision displaying various content feeds---posting vs. replying \& upvoting vs. downvoting.
	Extending on this, we derive further metrics defining community experience.
	Further, we show that the user influx and per user interactions can be modeled with a power law very well.

	\item
	We analyze user behavior \wrt{} lifetime and retention between different communities, which is very similar.
    However, daily user activity and churn scales with community size, whereas we observe differences in the interactions of users dropping out of the platform.

\end{itemize}

\afblock{Non-goals.}
Since the provided data resembles meta data only, without the posted content, we cannot provide a content analysis studying how and for what the Jodel network is used in the KSA.
We take the rare opportunity to use an operators view to empirically characterize the nationwide adoption of a new social media platform. 
We thereby refrain from deriving models to represent adoption processes in general given that our data covers KSA only.

An interesting aspect for future work is the comparison of adoption processed between different social networks.
This is a challenging question that is beyond the scope of our work.
The reason is that other networks like Facebook fundamentally differ in their design to the studied Jodel network (i.e., location-based vs. non-location-based, anonymous vs. non-anonymous).
Studying the influence of these fundamental design differences is in interesting study, yet a study that requires a dedicated and different study design and different data sets (e.g., comparable data sets from Facebook that are not available to us).
Thus, this question goes beyond the scope of our work, and we thus need to keep it for future work.
\begin{figure}[t]
	\centering
	\includegraphics[width=0.99\linewidth]{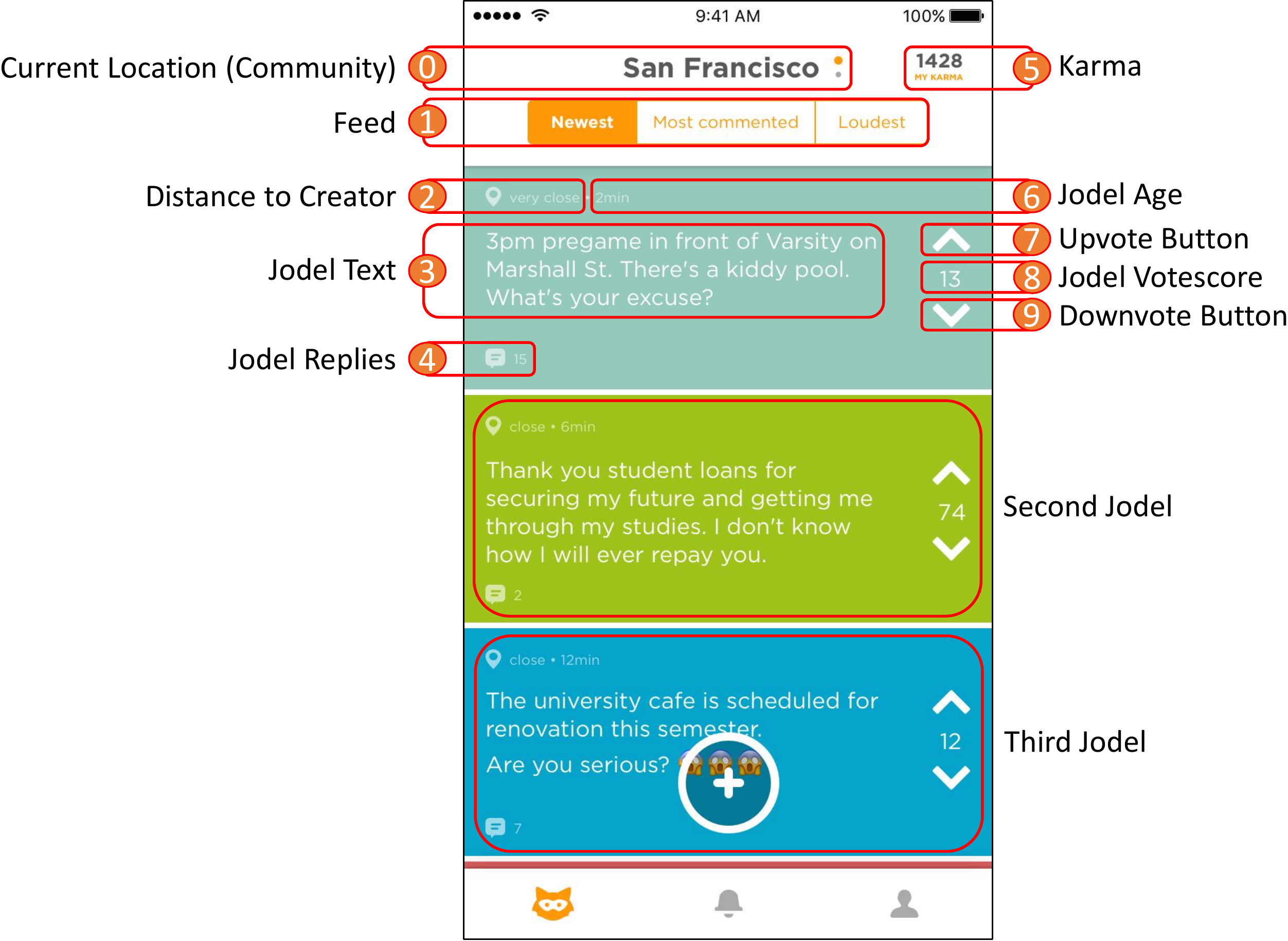}
	\caption{
		\textbf{Jodel iOS mobile application.}
	}
	\label{fig:jodelapp}
	\vspace{-1.5em}
\end{figure}

\vspace{-1.5em}
\section{Jodel - Anonymous Messaging App}
	\label{sec:jodelapp}

	Jodel\footnote{Jodel, German for yodeling, a form of singing or calling. ``Yudel'' \mbox{(\RL{يودل})} represents the adopted translation of Jodel to Arabic.}
	is a mobile-only messaging application which we show in Fig.~\ref{fig:jodelapp}.
	It is location-based and establishes local communities relative to the users' location \protect\encircle{0}.
	Within these communities, users can {\em anonymously} post both images\footnote{The ability to post videos and subscribe to channels was added after the end of our dataset.} and textual content of up to \numberNumCharactersPerJodel{} characters length \protect\encircle{3} (\ie{} microblogging) and reply to posts forming discussion threads \protect\encircle{4}.
	Posted content is referred to as ``Jodels'', colored randomly \protect\encircle{3}.
	Posts are only displayed to other users within close (up to 20km) geographic proximity \protect\encircle{2}.
	Further, all communication is {\em anonymous} to other users since no user handles or other user-related information are displayed.
	Only {\em within} a single discussion thread, users are enumerated and represented by an ascending number in their post order. %
	Up to 1500 threads are displayed to the users in three different feeds \protect\encircle{1}: i) \emph{recent} showing the most recent threads, ii) \emph{most discussed} showing the most discussed threads and iii) \emph{loudest} showing threads with the highest voting score (described next).

	Jodel employs a community-driven filtering and moderation scheme to avoid adverse content.
	For an anonymous messaging app, community moderation is a key success parameter to prevent harmful or abusive content.
	The recent downfall of the YikYak anonymous network (see~\cite{yikyakNyTimes}) highlighted that unsuccessfully preventing adverse content can seriously harm the network.
	In Jodel, content filtering relies on a distributed voting scheme in which every user can increase or decrease a post's vote score by up- (+1) \protect\encircle{7} or downvoting (-1) \protect\encircle{9}, \eg similar to StackOverflow.
	Posts reaching a cumulative vote score \protect\encircle{8} below a negative threshold \mbox{(\eg -5)} are not displayed anymore.
	Depending on the number of vote-contributions, this scheme filters out bad content while also potentially preferring mainstream content; or possibly opens up the possibility of DoS attacks from a multitude of computer-created accounts, which however is prevented with various operational measures, such as rate limiting.
	Further, every post can be flagged as abusive.
	Flagged content is displayed to voluntary, system-selected, community moderators that majority-vote to remove the particular post.

	To increase overall user engagement in terms of creating content and voting, the network applies lightweight gamification by awarding \emph{Karma} points \protect\encircle{5}.
	Karma points are collected by either actively voting or by receiving upvotes on posted content from others.
	Negative (down-) votes by other users decrease the Karma similar to posts that are removed from the system due to moderation.
	Achieved Karma scores are not displayed to other users but are a proxy for user activity and well-behavior (\eg to select community moderators).

\begin{figure*}[t]
	\centering
	\begin{subfigure}[t]{\linewidth}
			\includegraphics[width=.24\linewidth]{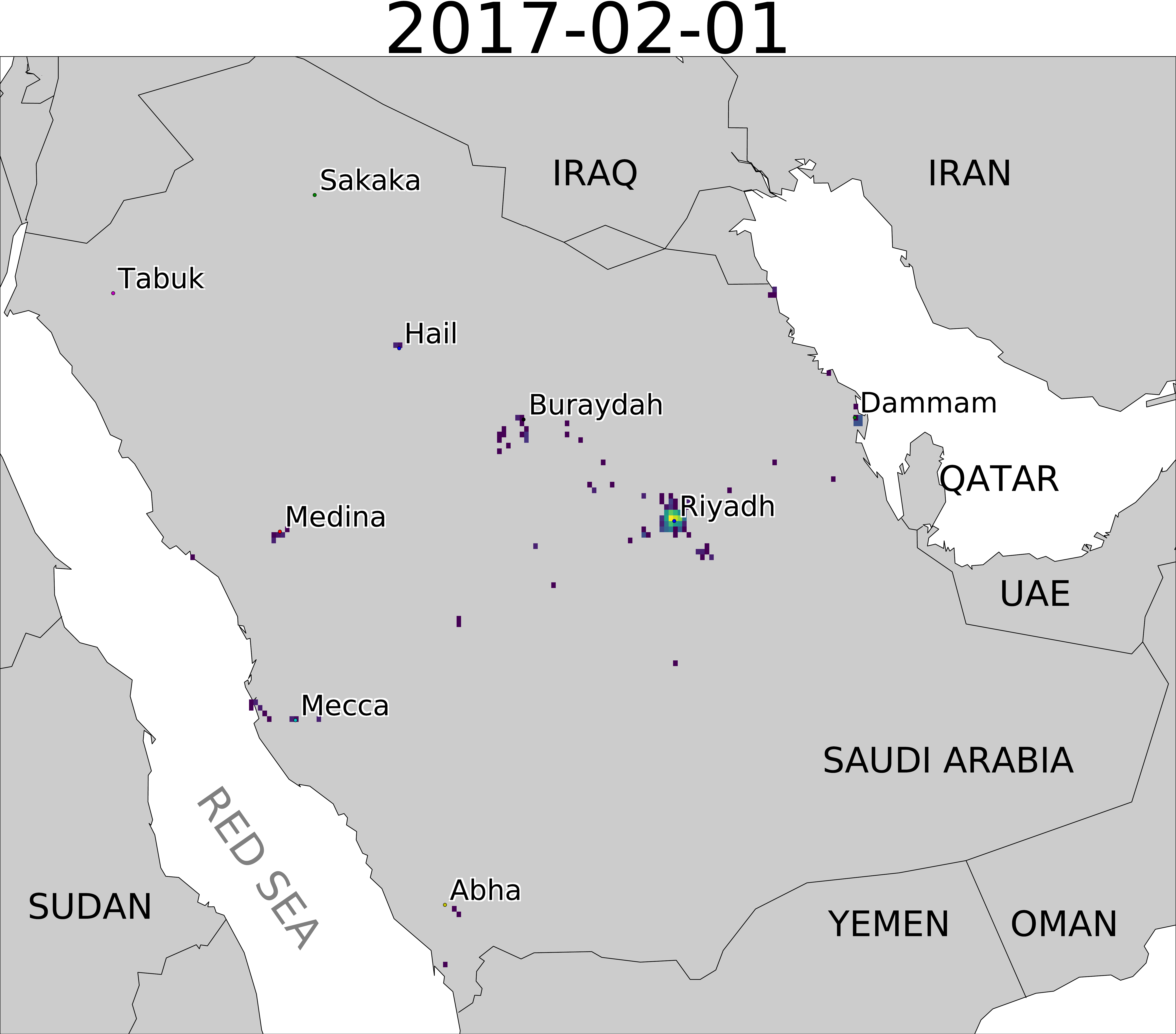}
			\includegraphics[width=.24\linewidth]{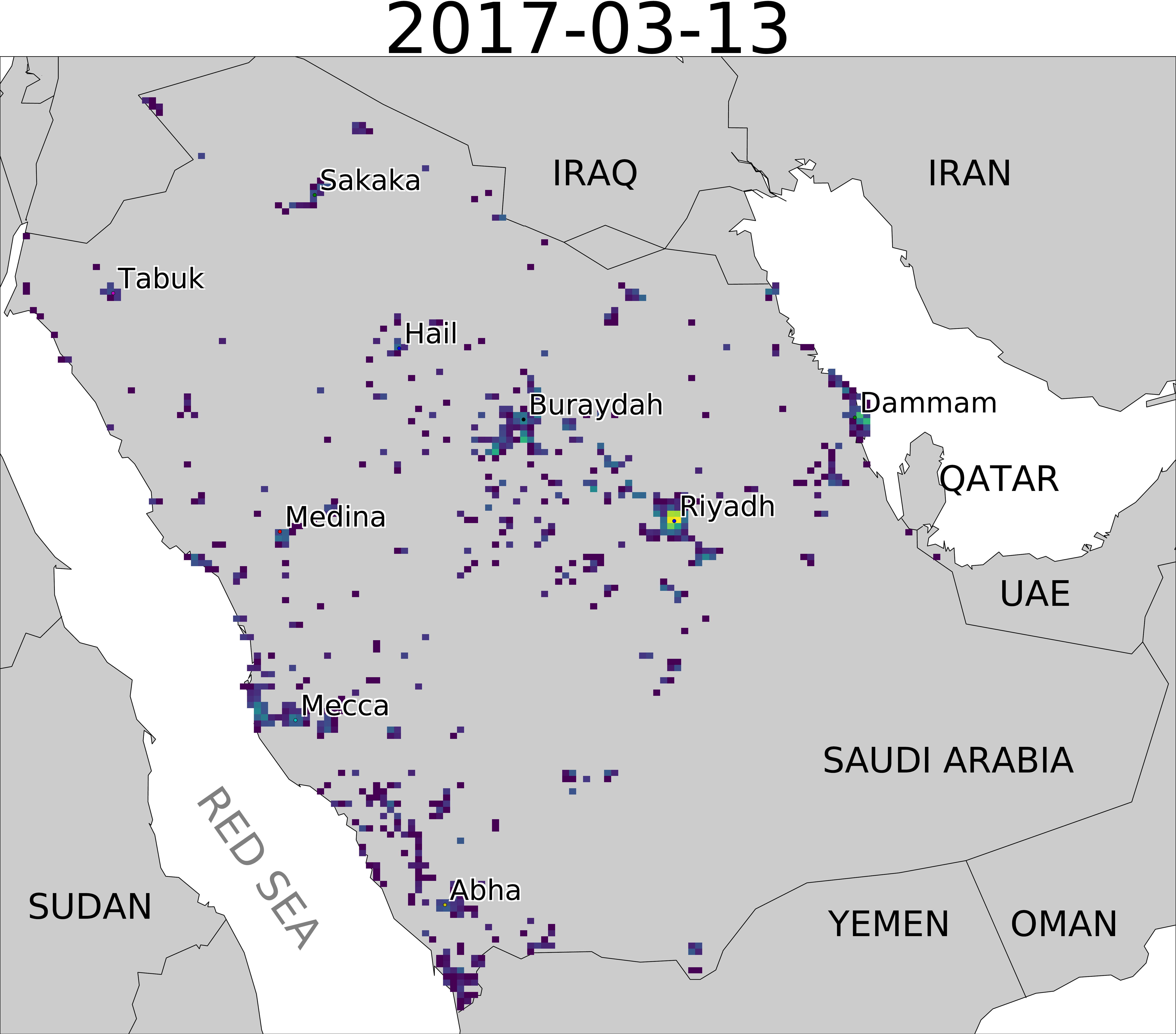}
			\includegraphics[width=.24\linewidth]{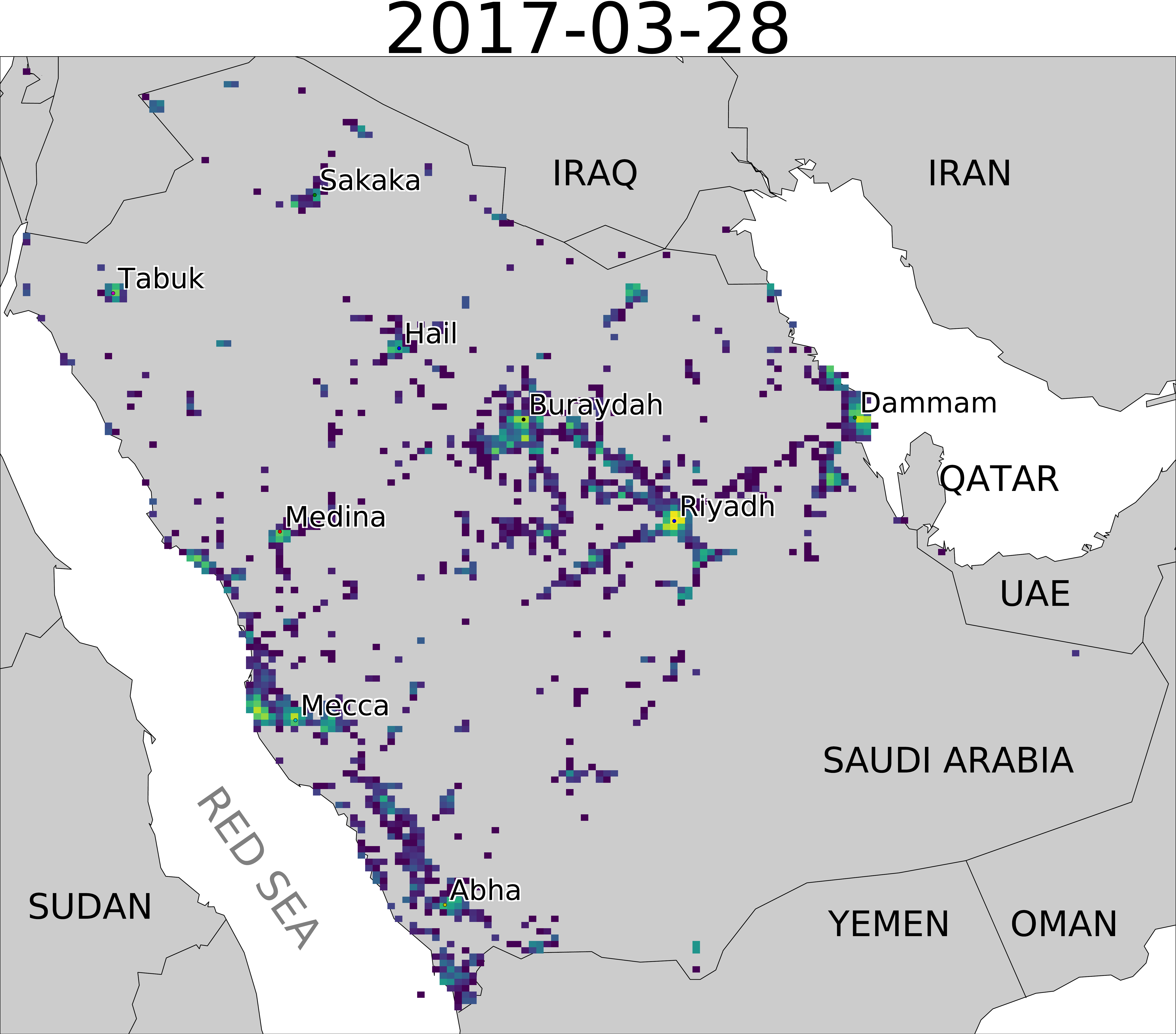}
			\includegraphics[width=.24\linewidth]{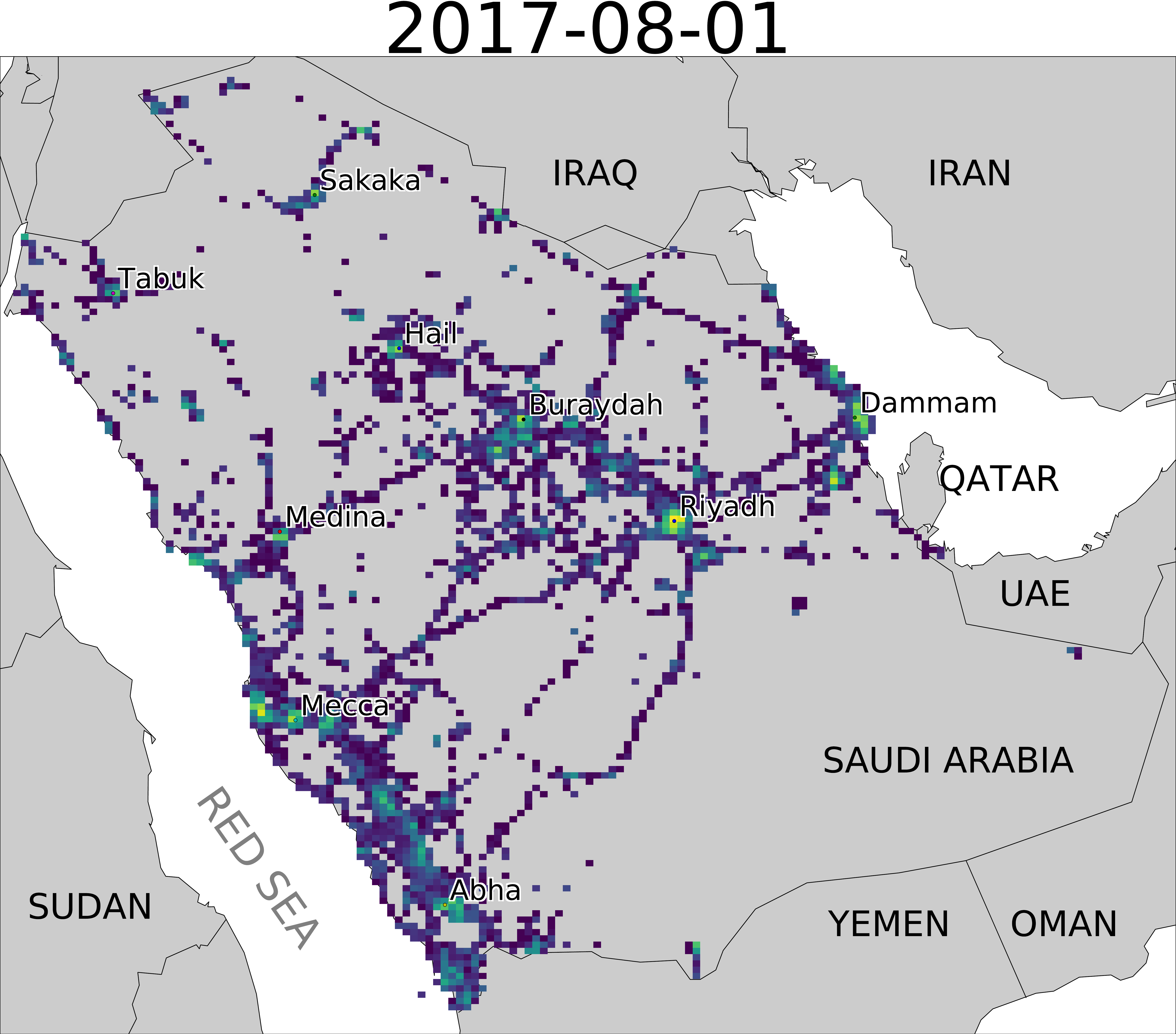}
			\includegraphics[width=0.01015\linewidth]{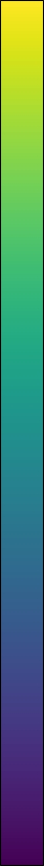}
		\caption{
			\textbf{Qualitative geospatial development of user interactions in the KSA in 2017.}
			This figure show snapshots in time (before the jump-start, at the beginning on March 13 and March 27 as well as Aug 1) across the country.
			The colored mesh depicts the number of system interactions log-normalized for each snapshot; lighter color describes higher activity.
			While the absolute amount of interactions in February is negligible, first users focus on the capital Riyadh.
		}
		\label{fig:interactions_map}
	\end{subfigure}

	\begin{subfigure}[t]{\linewidth}
		\includegraphics[width=\linewidth]{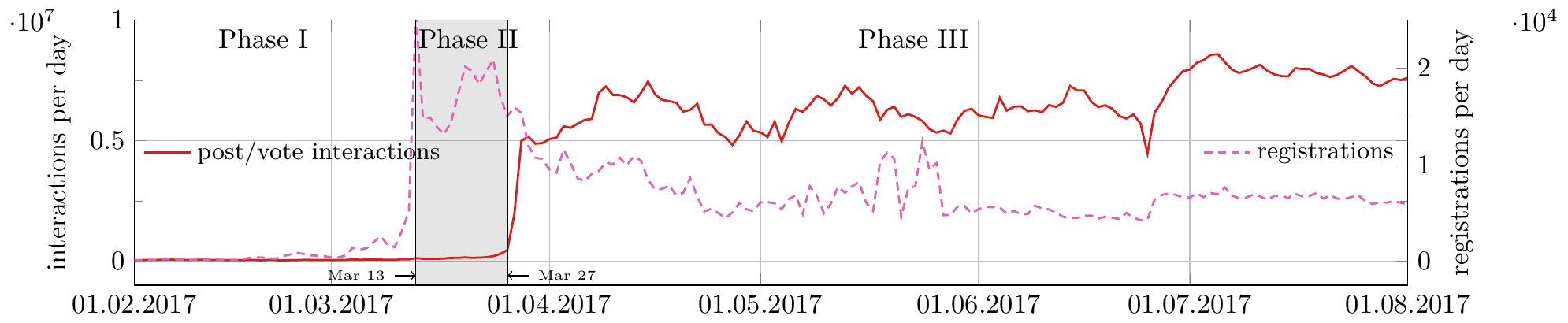}
		\caption{
			\textbf{Total user interactions in the KSA Feb to Aug 2017.}
			This figure describes the number of different user interactions with the system.
			The x-axis depicts the time, whereas the y-axis marks the amount (registrations right y-axis, other interactions left y-axis).
			New user registrations first peak on 13 March, whereas activity heavily explodes two weeks later.
			We are interested in who drives this jump-start and the subsequent user base development.
		}
		\label{fig:interactions_over_time}
	\end{subfigure}
	\caption{
		\textbf{Jodel usage adoption and development over time.}
	}
	\vspace{-1.5em}
\end{figure*}

\vspace{-1em}
\subsection{Dataset Description and Ethics}
	\label{sec:Dataset_Description_and_Statistics}
	The Jodel network operators provided us with {\em anonymized} data of their network, which we summarize in Table~\ref{tab:dataset}.
	The obtained data contains post and interaction \emph{metadata} created within the KSA only and spans multiple years from the beginning of the network up to August 2017.
	Note that Jodel is still being used within the KSA as of today.
	It is limited to metadata only without textual content and user records stripped and anonymized.
	The data enables us to cluster users by their anonymous ID.
	However, it does not contain the content of posted messages and thus cannot be used for content analysis.
	Further, it contains no personal information and cannot be used to personally identify users.
	We inform and synchronize with the Jodel operator on analysis we perform on their data.
	The structure of our available dataset includes three object categories: interactions, content, and users.

	\begin{itemize}[noitemsep,topsep=5pt,leftmargin=10pt]
	\item
		\textbf{Users} (about 1\,M records) only contain a user's accumulated Karma value and whether the user is blocked.
	\item
		An \textbf{interaction} (about 1\,B records) can be a \emph{registered}, \emph{post}, \emph{reply}, \emph{upvote}, \emph{downvote}, or \emph{flag}.
		Each interaction has a timestamp and a geohash.
		It is further linked to a user ID and a content ID.
	\item
		\textbf{Content} (about 500\,M records) may either be a new post, \ie starting a new thread, or a reply.
			This content includes a boolean flag whether it is text or an image (video added after our measurement ended).
		Each content record holds metadata about the accumulated voting-score and whether it has been blocked (\eg by moderation).
		Note that this content \emph{metadata} does not contain the posted text.
	\end{itemize}

	\begin{table}[t]
		\small
		\centering
		\begin{tabular}{|l|r|l|}
			\hline
			\textbf{Type}		& \textbf{\#Entries}	& \textbf{Description}\\ \hline\hline
			User				& 1,2\unit{M}			& User metadata\\\hline
			Content				& 469\unit{M}			& Content (posts, replies)\\\hline
			Interaction			& 1,155\unit{M}			& Interactions incl. user, geographic\\
								&						& position and type (post, reply, up-/\\
								&						& downvote, flag)\\\hline
		\end{tabular}
		\caption{
			\textbf{Dataset statistics.}
			The data ranges from the application start in late 2014 up to the beginning of August 2017. We find the first observation in the KSA in December 2014.
		}
		\label{tab:dataset}
		\vspace{-2em}
	\end{table}

	\afblock{Dataset limitations.}
	Our dataset only includes the users' {\em active} interactions with the system, where they contribute like registering, creating posts, replying, or voting.
	Thus, we cannot infer when or how much a user only {\em passively} participates---lurkers---who only consume content.
	Further, the vote interactions are always mapped to the date and geoposition of the respective content creation.
	This prevents us from making detailed analyses depending on the voting time or place.
	However, due to the vivid usage of the application (multiple posts/replies per minute), we generally consider votes to be executed on the same day as their respective content.
	Especially since posts are only accessible via the three different feeds, where they will only stay for a very limited time, casting votes long after the content creation is usually not possible.

	\vspace{-1em}
	\subsection{Partitioning the Communities}
	Given that communities are defined relative to the users' GPS position and thus volatile, we map activities to the nearby city to form stable references for analysis.
	In larger cities, our city-level aggregation can form larger communities than displayed to the users.
	
	For a subsequent analysis of the different communities in the remainder of this paper, we define interactions as active participation by either posting, replying, or voting on posts.
	To analyze communities, we apply an abstraction assigning interactions to nearby cities and sorting all communities into sets of cities \wrt{} their interaction volume.
	We manually investigated several cities across the whole country to check that the user behavior is not substantially different to the aggregates (not shown in detail).
	Since this is not the case, we base the analysis in this section on four groups of city aggregates, \emph{ranked} by their interaction volume: q0-25 represents the $<25\%$ cities with the lowest number of interactions while q75-100 represents the top 25\% cities by interactions.
	Note that community sizes are not equally distributed, but follow a power-law (not shown).
\vspace{-1.5em}
\section{Adoption of Jodel in the KSA}
	\label{sec:adoption}
	We start by analyzing the overall Jodel usage evolution in the Kingdom of Saudi Arabia (KSA).
	Thereby gaining first insights on our first research question: %
	how fast did the adoption in the KSA occur (\sref{sec:temporaladoption}) and do adoption patterns differ geographically (\sref{sec:geoadoption})?
	To answer it, we analyze how its users interact with the network in terms of registration and interactions (\ie{} content creation and voting events).

	A user becomes part of the Jodel network via a device registration event.
	That is, whenever a newly recognized device starts the Jodel application, the system automatically assigns the device a new user account in the background.
	We show the geospatial development of Jodel within the KSA in Figure~\ref{fig:interactions_map}.
	The figure shows the app interaction activity as heatmaps on a per snapshot logarithmic scale for four days in 2017: February 1, March 13, March 28 and August 1 (left to right).
	A lighter/darker color indicates a higher/lower amount of activity, respectively.
	At the beginning of 2017, the capital Riyadh was practically the only city where Jodel was used, while the adoption swept over all major cities later on.
	We next begin with a detailed study of the temporal phases of Jodel adoption in the KSA (\sref{sec:temporaladoption}) and then study the adoption pattern of the different Jodel communities (\sref{sec:geoadoption}).

    \vspace{-1em}
    \subsection{Temporal Adoption}
	\label{sec:temporaladoption}
	\afblock{Phase I: Early Inception (2016).}
	A first peak in usage and registrations can be traced back to August 3, 2016 (not shown). %
	With negligible activity before in the KSA, the Jodel adoption grew by 140 and 170 users on two consecutive days.
	Afterwards, it experienced a small, but steadily increasing influx of new users.
	Due to the growing community, also the number of system interactions increased from $1\text{k}$ and $5\text{k}$ posts/replies on the first two days to more than $15\text{k}/\text{day}$ posts and replies throughout August 2016; the number of up- and downvotes evolved similarly.
	This early adoption coincides with an update of the similar YikYak application which introduces user handles and profiles, and thus abandoned anonymous posting capabilities (see~\cite{yikyakchange}).
	This finding {\em may} suggest that users switched to Jodel to keep-up posting anonymously.
	While this marks the birth of Jodel in the KSA, its widespread usage started months later.
	\label{sec:sudden_growth_march}
	\afblock{Phase II: Sudden Growth in March.}
	On March 13, 2017 the number of new user registrations and on March 27th, the number of messages posted to Jodel within the Kingdom of Saudia Arabia (KSA) increased almost 100-fold over the previous weeks, and continued to increase from there over the following weeks.
	The increasing app usage is highlighted in Figure~\ref{fig:interactions_over_time} showing the number of daily user interactions (y-axis) with the network by their type over time (x-axis).
	We omit interactions before February 2017 since there was only little usage within the KSA that is not directly visible in the plot.
	The registrations suddenly peaked on March 13 at \numberMarchPeakRegistrations{} new registrations and then decreased afterwards.
	The number of new registrations later settles at about $7.5\text{k}/\text{day}$ beginning in June.
	We define this sudden growth in both user registrations and the actual system usage in March as the beginning of the widespread adoption of Jodel within the KSA.
	We call this sudden adoption happening \emph{jump-start}.

	\afblock{Signs of external triggers.}
	This observation opens the question on what triggered the huge influx of new user registrations in March 2017.
	Since the design of the Jodel app inherently limits the ability of users to only communicate with others in close proximity, the large influx of new users at a country-wide basis was likely triggered \emph{externally} rather than originating from internal growth.
	One would suspect that such a jump-start has its origin in either marketing or promotional activities---or by mentions of public figures.
	Knowing that the Jodel company did \emph{not} launch any advertising in this region, the origin must be driven by users, advertising Jodel via external platforms.
	Since the Jodel user base is anonymous, we cannot provide ground truth information by interviewing early adaptors on their motivation to start using Jodel.
	However, the sudden peak in March is correlated to increasing attention to the Jodel app on other online platforms.
	Examples include search activity for the Arabic term ``Yudel'' \mbox{(\RL{يودل})}~\cite{GoogleTrends}.

\begin{figure}[t]
    \centering
    \includegraphics[width=1.\linewidth,frame]{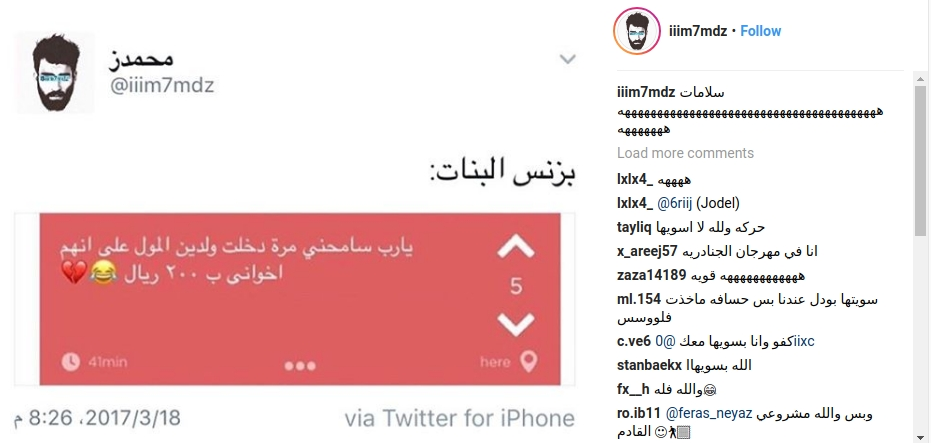}
    \caption{
        \textbf{Example Instagram post sharing a Jodel post} (red part).
        An \emph{entertaining} story about a girl accompanying two stranger guys as their sister to enter the mall for money (which they are not allowed to without family).
        The Instagram user comments ``Girls business'' and gathered 5,840 likes. %
		We have also interviewed this influencer. %
    }
    \label{fig:twitter}
	\vspace{-1.5em}
\end{figure}

	To look for external triggers, we manually inspected the social media platforms Twitter and Instagram, given their popularity in Arabic speaking countries (see~\cite{dennis2016media}).
	This way, we identified 15 KSA-based influencers (\ie social media users followed by a large number of users) who have shared \emph{funny} content originally posted on Jodel on their social media accounts~\cite{twitter3,twitter4,instagram1,instagram2,instagram3,instagram4} within the time frame when the registrations started to peak.
	Figure~\ref{fig:twitter} is just one of many examples in which the user \emph{\cite{twitter4}} (694k followers on Twitter and 3.5M followers on Instagram as of June 2020) shared Jodel content within these two weeks of the registration jump-start.

	For gaining a better initial insight on who triggered whom, we then contact these 15 identified influencers and asked about their motivation for using Jodel and how they got to know it.
	7 users replied to our inquiry (including \emph{iiim7mdz} shown in Figure~\ref{fig:twitter}).
	While there are supposedly several reasons to use Jodel, a major benefit mentioned multiple times was to get in touch with new \emph{local} people and especially women in an \emph{innocent} way.
	Another opinion stated that using Jodel just became very mainstream, a ``boon'' \mbox{(\RL{هبة})}.
	From these interviews, we hypothesize that the root cause of this jump-start indeed seems to be a wave of postings through the Arabic speaking social media landscape.
	Our data, however, does not enable us to pinpoint the very first key events nor to ultimately clarify why this jump-start happened.

	\afblock{Delayed interaction startup.}
	While our investigations {\em may} explain the sudden increase in registrations, they do not explain the delayed interaction with the system (see Figure~\ref{fig:interactions_over_time}).
	By manually analyzing social media posts published during the two weeks of the registration peak, we observed that users at first did not know how to use the application by asking how it works.
	This confusion on how Jodel works can explain why the peak in registration is not directly followed in heavy usage alike. %

	\afblock{Phase III: nation-wide establishment.}
	The last phase is characterized by a nation-wide usage and a continuous influx of new users and increasing number of interactions with the Jodel platform. 
	It marks the apps' establishment. 

	\takeaway{The adoption of Jodel in the KSA began in Riyadh (capital) and happened in phases, most notably a phase of sudden adoption triggered by external users.}

\begin{figure}[t]
    \centering
    \begin{subfigure}[t]{.475\linewidth}
		\centering
        \includegraphics[width=1\linewidth]{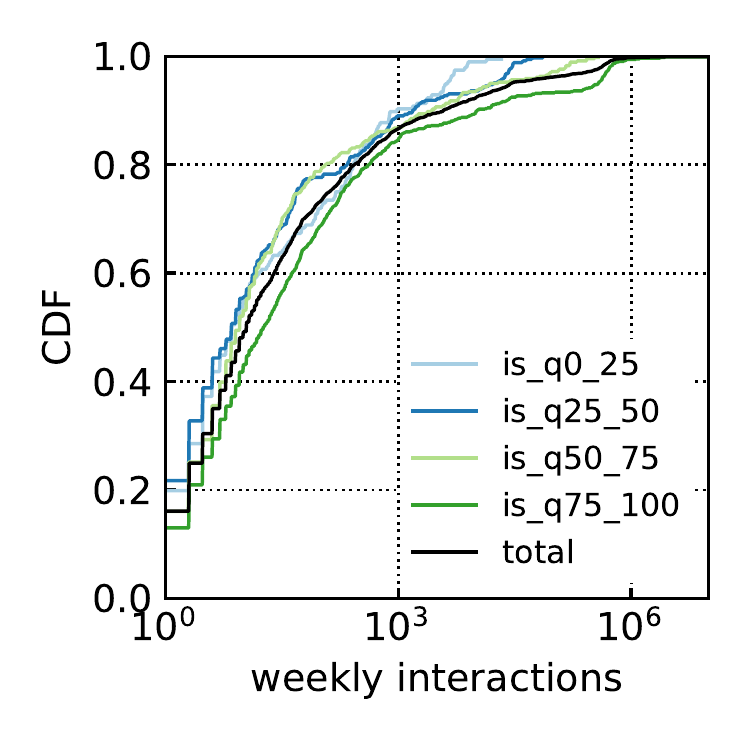}
        \caption{
            \textbf{Phase I-II.}
        }
        \label{fig:cdf_interactions__phase1_2}
    \end{subfigure}
    \quad
	\begin{subfigure}[t]{.475\linewidth}
		\centering
        \includegraphics[width=1\linewidth]{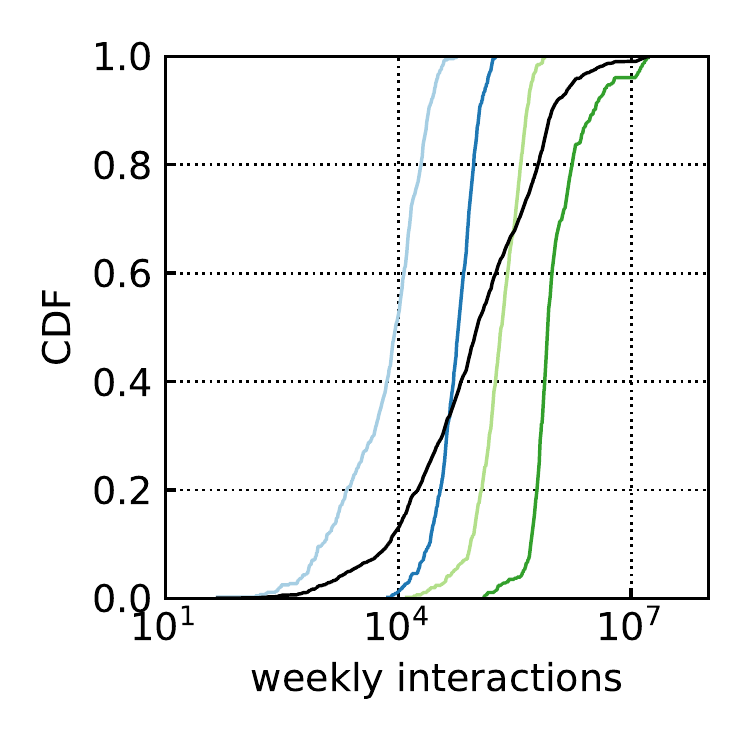}
        \caption{
            \textbf{Phase III.}
        }
        \label{fig:cdf_interactions__phase3}
    \end{subfigure}
    \caption{
        \textbf{Community-Quantile Interaction aggregates by week.}.
		We overse strong power-law distributions for community interactions.
        \mbox{\emph{(a)}} The amount of weekly interactions remains qualitatively equal between later determined community sizes.
        \mbox{\emph{(b)}} As per design, with stabilized community interactions, the weekly interactions tear apart in their quantiles.
    }
	\label{fig:cdf_interactions}
	\vspace{-1.5em}
\end{figure}

\vspace{-1em}
\subsection{Geographic Adoption \& Growth}
\label{sec:geoadoption}
We next answer the question if the observed 3-phase adoption pattern occurred similarly at all regions.
That is, do all the independent local Jodel communities show the same adoption pattern as in Figure~\ref{fig:interactions_over_time} (\eg{} the heavy user influx in March 2017)?
This location-based nature yields the formation of hundreds of independent communities throughout a country---without any country-wide communication.
Thus, differences in the adoption pattern between communities could be expected.

\afblock{Community Interaction Volume.}
We will base our subsequent analyzes per design an overall community interactions through partitioning them in quantiles.
However, to better grasp the distribution of interactions between these groups, we present Cumulative Distribution Functions (CDFs) of weekly figures in two different time periods:
\emph{Left)} Phase I-II in Figure~\ref{fig:cdf_interactions__phase1_2}, and \emph{Right)} Phase III in Figure~\ref{fig:cdf_interactions__phase3}.
Both CDFs show the city quantile subsets and the total CDF in comparison.
While the distributions are qualitatively similar, we do not observe major variations, \ie{} though interactions follow are power-law across communities.
As per design, the CDFs tear apart within Phase III.

\takeaway{
	Jodel remained largely unknown within the KSA up to mid March in 2017.
	With high confidence, we suspect external Social Media having triggered a sudden growth in popularity afterwards.
}

\vspace{-1em}
\section{Platform Interactions}
\label{sec:community}
The primary function of the Jodel app and social media in general revolves around user interactions.
In the case of Jodel, framed into location, this is restricted to anonymous content sharing, communication and dis- or liking contents.
We next discuss the dynamics of these platform interactions.

\begin{figure*}[t]
    \centering
    \begin{subfigure}[t]{.32\linewidth}
		\centering
        \includegraphics[width=1\linewidth]{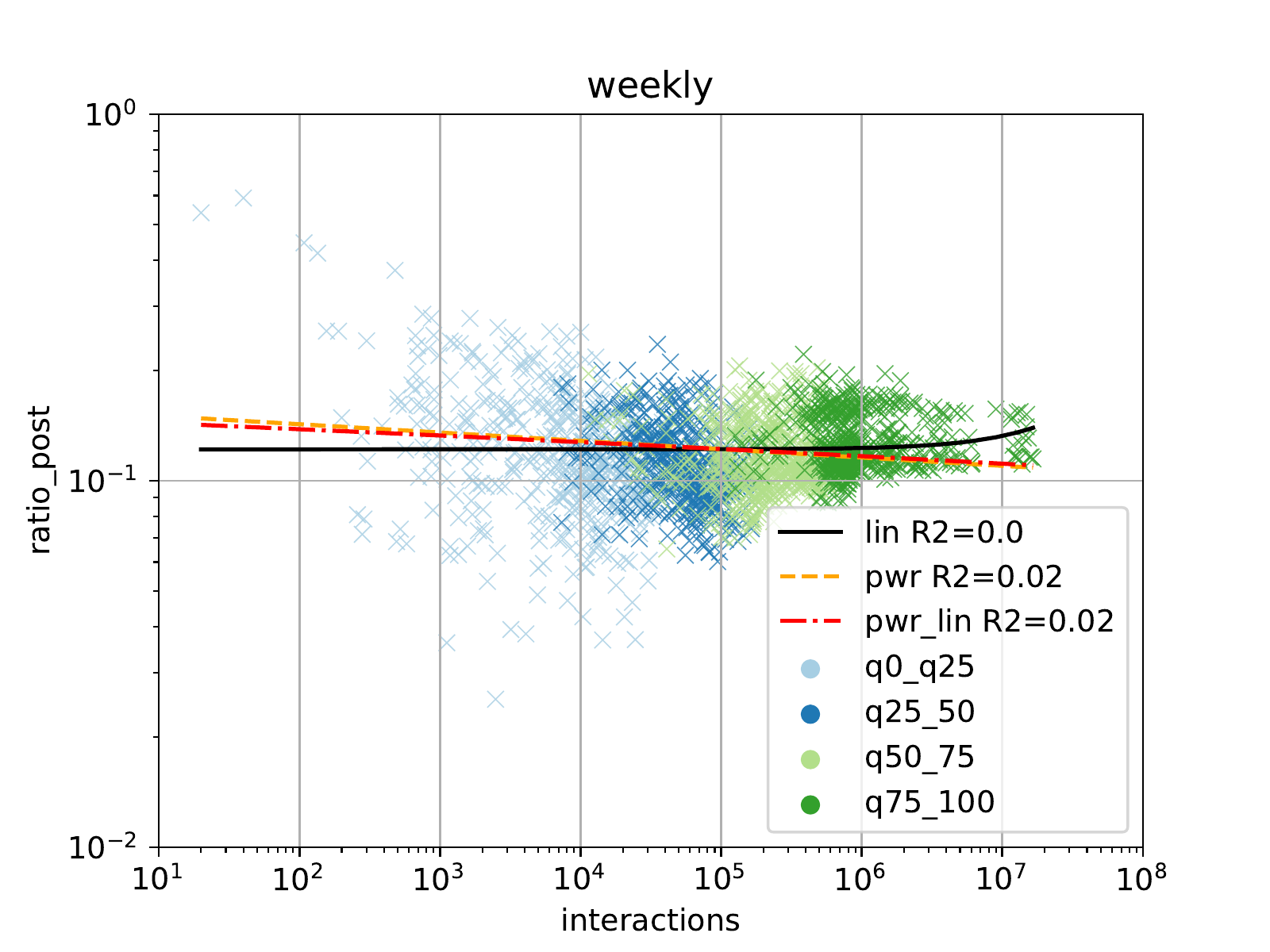}
        \caption{
            \textbf{\#Posts $\times$ \mbox{(\#Posts+\#Replies)}.}
        }
        \label{fig:postratio_scatter}
    \end{subfigure}
    \begin{subfigure}[t]{.32\linewidth}
			\centering
			\includegraphics[width=1\linewidth]{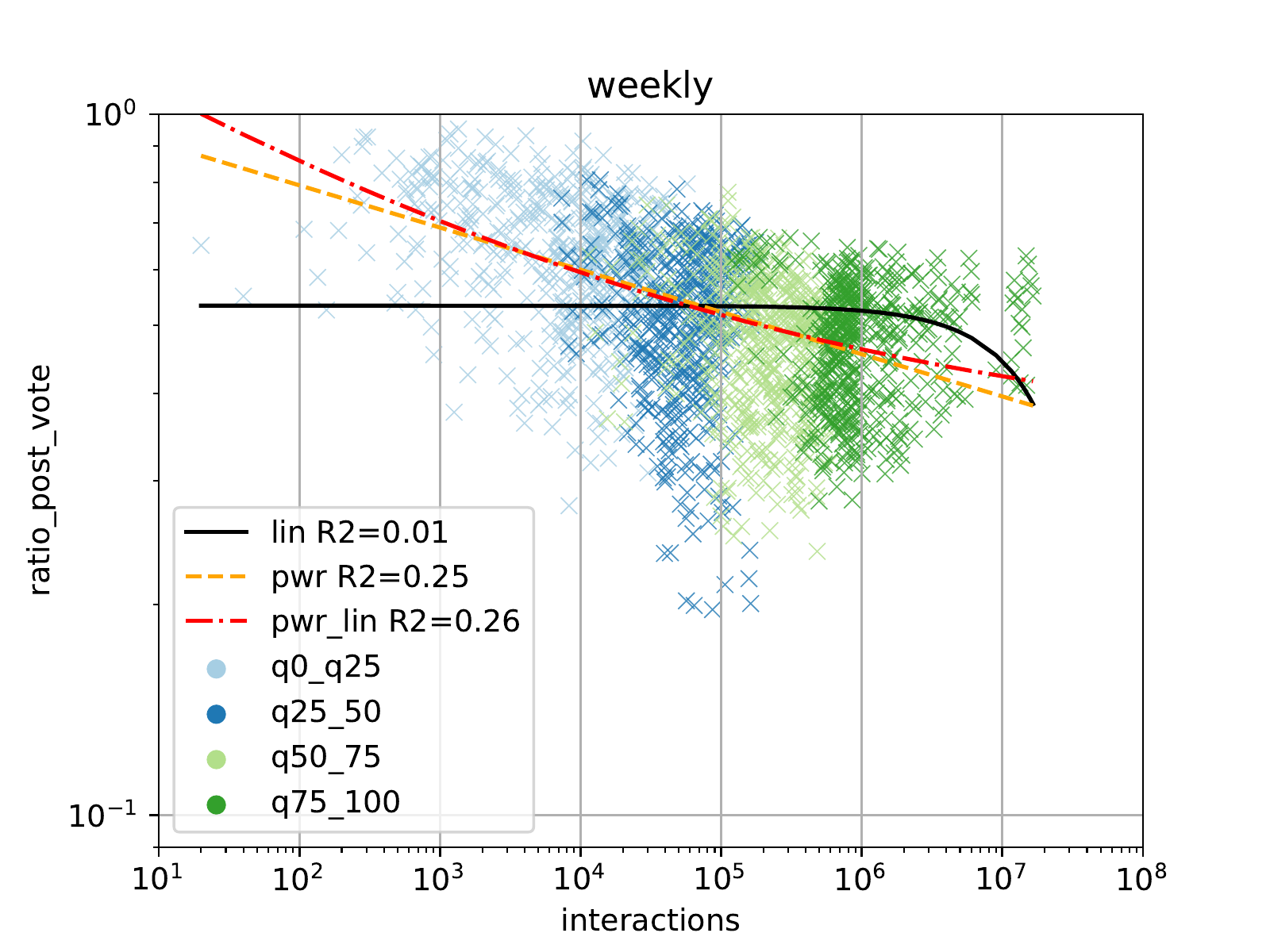}
            \caption{
            \textbf{\#Content $\times$ \mbox{(\#Content+\#Votes)}.}
        }
        \label{fig:votes_posts_scatter}
    \end{subfigure}
    \begin{subfigure}[t]{.32\linewidth}
		\centering
			\includegraphics[width=1\linewidth]{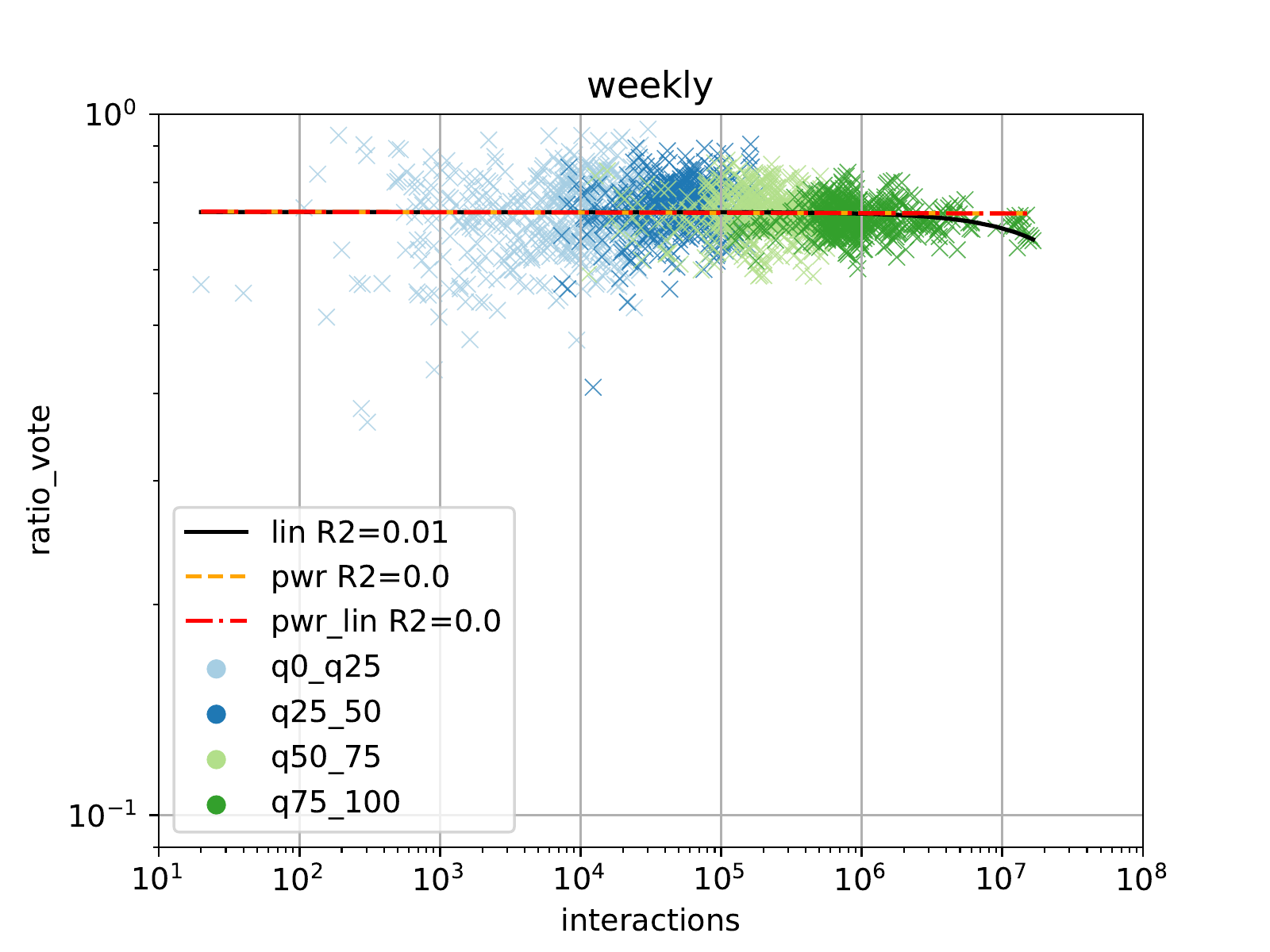}
		\caption{
			\textbf{Happyratio (\#Up- $\times$ \#Votes), \mbox{rec}.}
		}
		\label{fig:votes_scatter}
    \end{subfigure}

    \caption{
        \textbf{User interaction ratios by city size.}
        \emph{(a)}
        This plot depicts \#posts to total content by city size.
        Sarting new threads is less popular in general with a slight downwards trend with city size on average.
        \emph{(b)}
        This plot depicts the ratio of content creation to votes to by city size.
        With increasing community sizes, the amount of created content to votes converges towards equal popularity.
        \emph{(c)}
        This plot shows \#upvotes to total votes by city size.
        Overall, positive votes are dominating.
        While being noisy for smaller cities, bigger communities tend to be more critical in their votes.
    }
    \vspace*{-1.5em}
	\label{fig:community_interactions_metrics}
\end{figure*}

\subsection{Interaction Dynamics}
\label{sec:interaction_dynamics}

Since the Jodel app design provides three different content feeds (\emph{most recent}, \emph{most discussed}, and \emph{loudest}), we are interested in relations between interaction amounts---compared to our partitioning driver community-size. 
Thus, we next set the stage and provide an overview of how the total platform interactions types, \eg{} posting or upvoting, interlink to each other.
In Figure~\ref{fig:community_interactions_metrics} we provide three different scatter plots of weekly averages over interactions per community (x-axis), colored by their quantile.
We added the following fitting curves and their fitting R2 score to ease interpretation, which will also be used throughout this paper in subsequent evaluations:
\begin{align*}
	\text{linear} 	   	     & & \text{lin}       &  & = & & a & & + & & b & & x&			\\
	\text{power law}		 & & \text{pwr}       &  & = & &   & &   & & b & & x&^c		\\
	\text{shifted power-law} & & \text{pwr\_lin}  &  & = & & a & & + & & b & & x&^c
\end{align*}

\afblock{Creating Content.}
First, we measure the postratio (y-axis) denoting the amount of threads (posts) compared to the total amount of content (posts and thread replies) in Figure~\ref{fig:postratio_scatter}, that is a postratio converging towards 1 is post dominated, whereas reply dominated converging to 0 exemplified next.
Together with conversationness (\sref{sec:structural_implications}), we argue that postratio is a vital key indicator in favor of communication (\texttt{n} to \texttt{m}) instead of predominantly shouting content out (\texttt{1} to \texttt{n}).
Due to rather steady-state community behavior as discussed in \sref{sec:geoadoption}, as per design, defined quantiles (color) fit nicely the weekly amount of interactions by community in distinct regimes, \ie{} there is an apparent gradient between data points across quantiles.
For largest communities, postratios vary from $0.1$ to $0.2$, \ie{} only 1 in 10 or 2 in 20 total content contributions resembles a new thread, whereas, with fewer interactions, values gradually vary stronger, \eg{} from $0.06$ to $0.23$ for q25-50, and even more so for q0-25.
Nonetheless, we identify a very homogeneous distribution of threads and replies across all independent communities at a postratio$\pm$std of $0.122 \pm 0.038$ on average, \ie{} there are about 12 threads in 100 content contributions.
The imbalance between threads and considerably more replies heavily influences user experience \wrt{} content perception due to replies being one of three in-app content feeds, \emph{most discussed}, as described in \sref{sec:jodelapp}.

\begin{figure*}
	\centering
    \begin{subfigure}{.31\linewidth}
        \centering
        \includegraphics[width=1\linewidth]{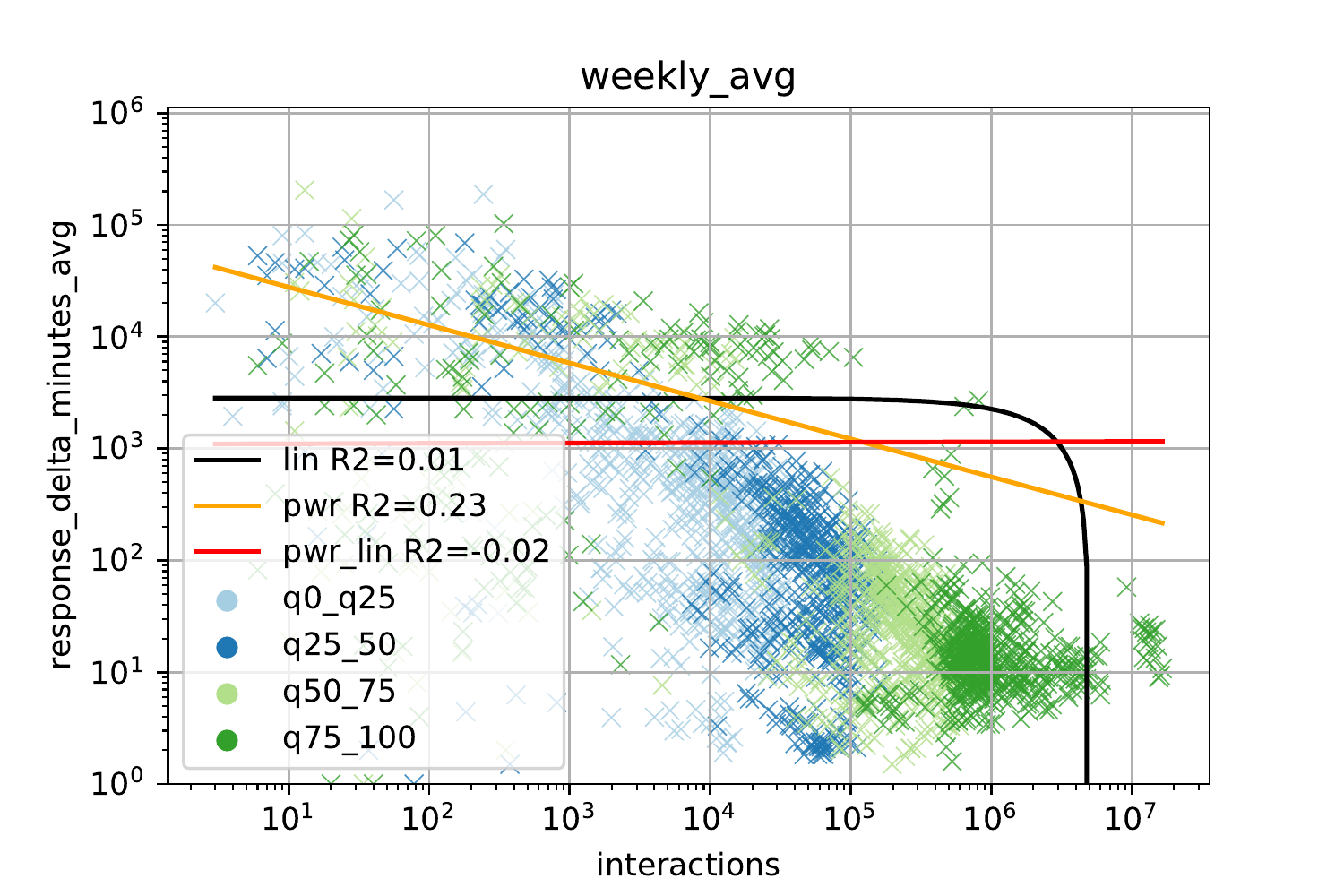}
        \subcaption{
            \textbf{Interactions $\times$ Reply Times}
        }
        \label{fig:time_to_reply}
    \end{subfigure}
    \quad
    \begin{subfigure}{.31\linewidth}
        \centering
        \includegraphics[width=1\linewidth]{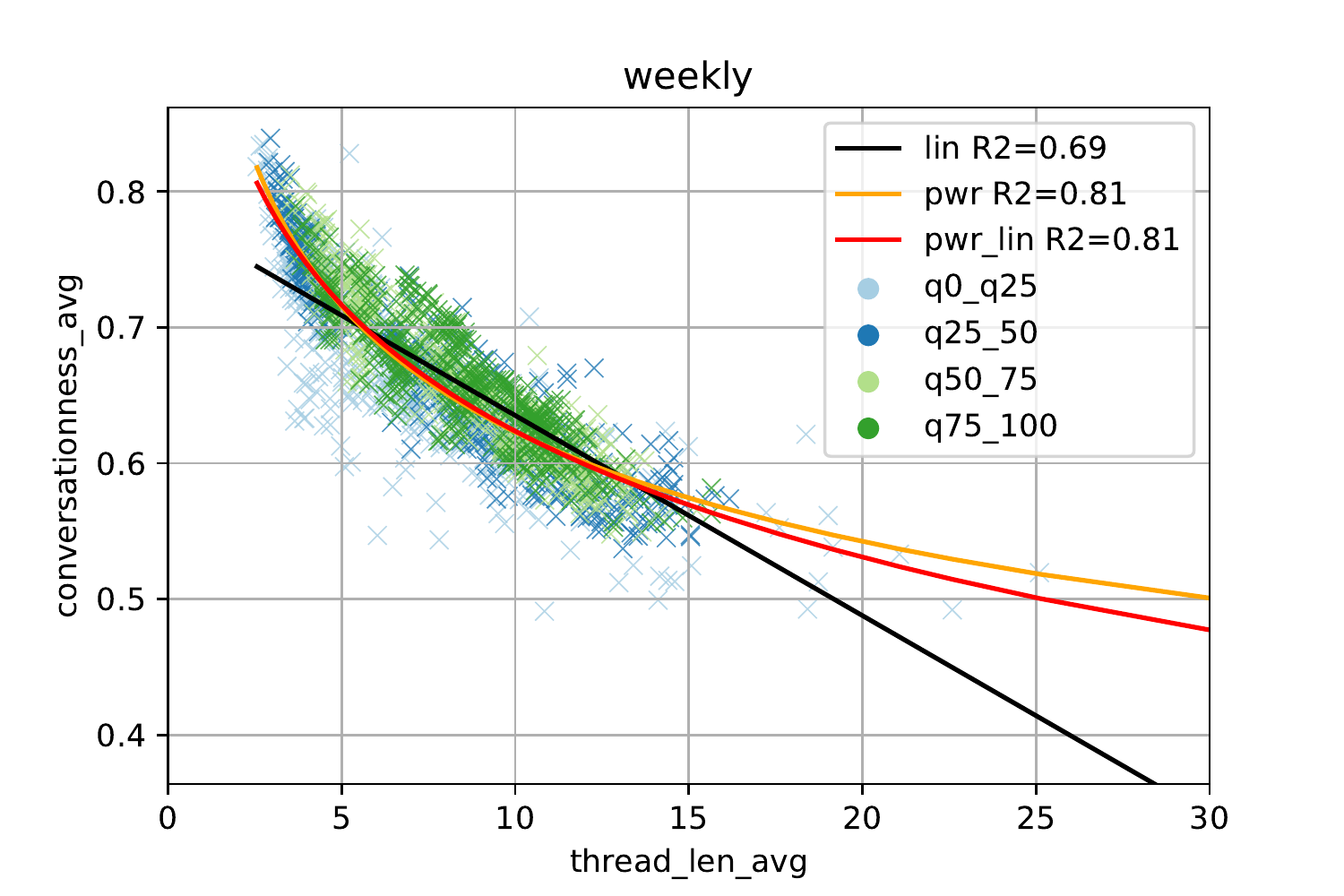}
        \subcaption{
            \textbf{Thread Len $\times$ Conversationness.}
        }
        \label{fig:conversationness}
    \end{subfigure}
    \quad
    \begin{subfigure}{.27\linewidth}
        \centering
        \includegraphics[width=1\linewidth]{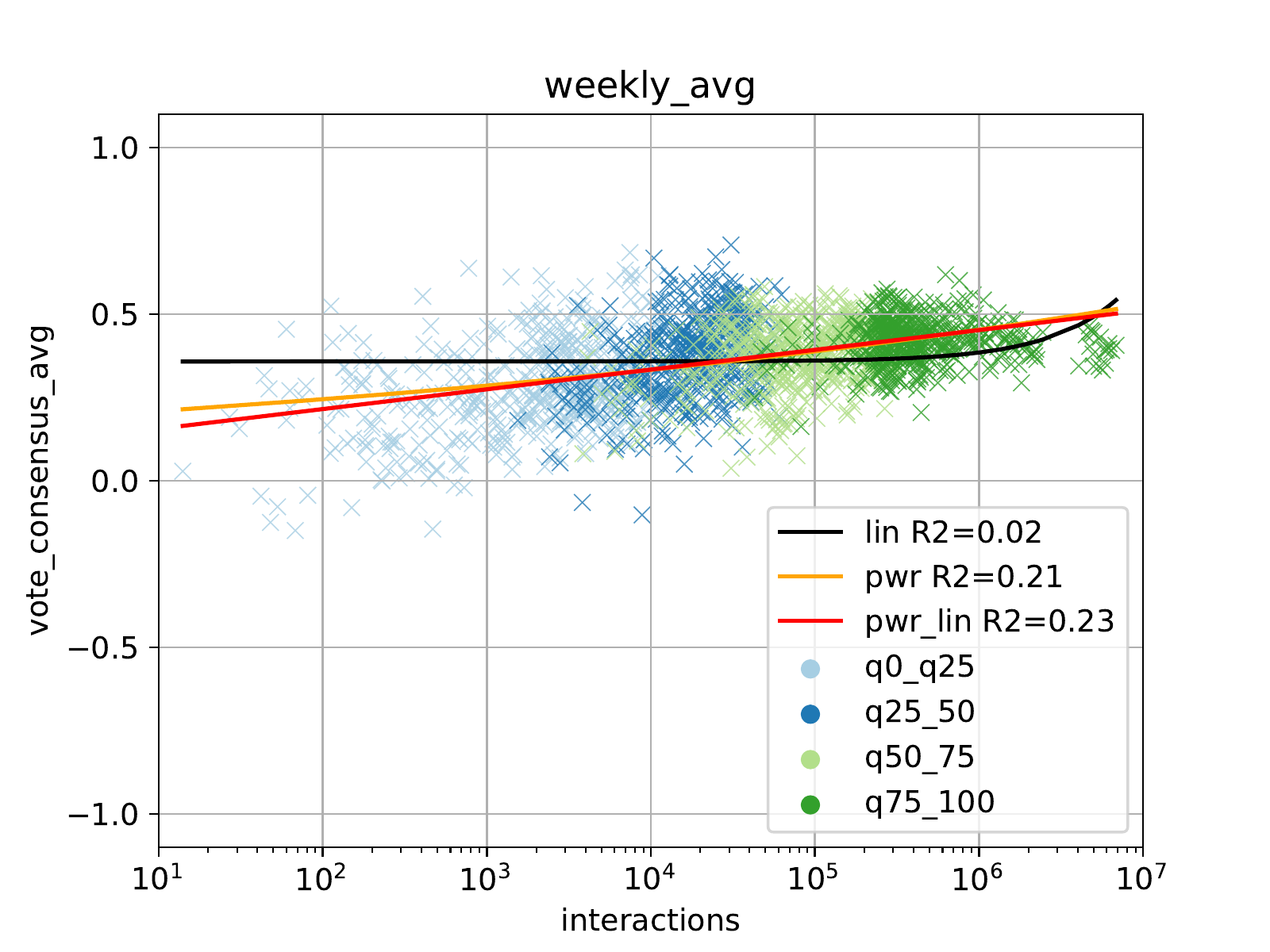}
        \subcaption{
            \textbf{Interactions $\times$ Consensus.}
        }
        \label{fig:vote_consensus}
    \end{subfigure}
    \caption{
        \emph{(a)}
        The average time until a post receives a response average per city on a weekly aggregate.
        The distributions are heavily scattered, while we observe a natural trend towards lower response times with increasing interactions.
        \emph{(b)}
        The average conversationness across communities on a weekly aggregate.
        Larger thread length correspond to a decreased conversationness indicating that discussions are of mixed nature in number of participants.
        \emph{(c)}
        While the average vote consensus varies for smaller communities, it converges towards values of 0.5, \ie{} 75\% of total votes are upvotes.
    }
    \vspace{-1.5em}
\end{figure*}

\afblock{Content Frequency.}
On weekly averages, we observe more variation for smaller communities.
Nontheless,
\emph{i)} the postratio (posts/total content) is mostly identical across all communities, 
and \emph{ii)} the content to total interaction ratio is within arguably similar regimes with a slight downwards trend with community size,
the amount of new posts per timeframe qualitatively follows the interaction distributions as shown in Figure~\ref{fig:cdf_interactions__phase3}.
Thus, the resulting frequency of new posts within the local \emph{most recent} app-feed heavily increases with community activity, that can be modeled very well with  a shifted power-law with a R2 score of 0.98.
Note that these averages do not account for a day/night cycle: rendering the update frequency higher at night.
To put this into perspective within Phase III:
In the largest community Riyadh, we find about 85 new posts per minute.
The second-largest community Jeddah only experiences 29 new threads per minute.
However, the most recent feed for Q0-25 communities only gets updated every five minutes.

\afblock{Content \& Voting.}
While content creation forms the basis of the independent communities, only distributed voting brings them to life---making content disappear and favoring mainstream contributions.
To bring both interaction types in relation, we scatter plot the ratio of content to votes (y-axis) in Figure~\ref{fig:votes_posts_scatter}, that is values converging towards 1 (0) are post (vote) post dominated.
Again, as per design, we observe a color gradient along the x-axis as expected.
Though on average, content dominates total interactions on average at a ratio$\pm$std of $0.53 \pm 0.13$, from our curve fittings, a linearly shifted power law describes a ratio downwards trend best, \ie{} we interpret provided figures as a shift towards higher amounts in replies compared to threads with increasing interactions.
Albeit applying a low-pass filter by averaging, we find further confidence of this trend within quantile averages$\pm$std: \{q0-25: $0.64 \pm 0.14$, q25-50: $0.52 \pm 0.12$, q50-75: $0.48 \pm 0.10$, q75-100: $0.48 \pm 0.09$\}.
Overall, while being very noisy, we observe that both interaction types are equally popular across communities on average rendering any other platform interaction being either contributing content or voting.
However, there is a subtle gradient towards a post-domination regime for smaller communities.
Although this ratio is not inflicted to any feed, we argue that it provides a first key insight of community capacity for the applied distributed moderation scheme; for now, we want to skip this discussion until vote consensus in \sref{sec:structural_implications}.

\afblock{Voting.}
Voting is an essential feature for any social media platform to promote popular content---and is leveraged for distributed moderation.
Shedding light on votes, we describe in Figure~\ref{fig:votes_scatter} the happyratio (y-axis) as the fraction between upvotes and total votes; \ie{} values converging to 0 (1) are downvote (upvote) dominated.
As per design, we again observe a color gradient along weekly interactions (x-axis).
On average, the overall happyratio$\pm$std remains quite positive at a score of $0.72 \pm 0.06$ across all communities.
While we observe slightly higher standard deviations within smaller communities of stddev\textsubscript{q0-25}=$.09$, this value decreases for larger communities stddev\textsubscript{q75-100}=$.04$.
Irrespective of community size, 72\% of overall casted votes are upvotes, and therefore positive.
We argue that this metric is a key indicator to community sentiment; though downvotes are an integral part of the moderation system.
As such, a dominance in positive votes not only keeps a community running, but also influences user experience \wrt{} content perception through the \emph{loudest} content feed (cf. \sref{sec:jodelapp}).

\takeaway{
    \emph{i)} 
    More or less \emph{invariant} to community size, discussion are by far more popular (5-fold) than creating new threads, promoting the \emph{most discussed} app feed.
    \emph{ii)} 
    Though replies are dominating content, this results in a \emph{scaling effect}, that is represented by a power-law very well; updates of the \emph{most recent} app feed range from once per 5 minutes for q0-25 up to 85 posts per minute for q75-100 on average.
    \emph{iii)} 
    With higher variation results in a controversial picture, voting and creating content are equally popular throughout all communities on average with a downwards trends across increasing interactions.
	\emph{iv)} 
    \emph{Invariant} to community size, we find a strong bias towards positive votes a happyratio of 0.72 on average promoting higher scores within the \emph{most popular} app feed.
}

\vspace{-1em}
\subsection{Platform Implications}
\label{sec:structural_implications}

Having set the stage in providing first insights to overall content- and vote interactions, we next keep our community perspective and discuss key structural implications.
Due to the network living off of new started threads (posts) and especially discussion within these threads, we next focus on thread response times and conversationness that measures discussion participant homogeneity.
Further, as the platform relies on user voting for content steering, we also investigate voting consensus measures.

\afblock{Community Response Time.}
For measuring response times, as a simple proxy, we restrict our evaluation to the response time of the very first reply to a thread.
Note that due our dataset using post timestamps for votes as well, vote interactions do not allow for this evaluation (\sref{sec:Dataset_Description_and_Statistics}).

In Figure~\ref{fig:time_to_reply}, we show a scatter plot time until a first response to a post occurs over community interactions.
While the log scaled x-axis shows denotes the weekly interactions, the log scaled y-axis denotes the time to a first answer measured in minutes.
The shading represents the city quantiles; Lines denote applied curve-fittings.

While we observe largely noisy distributions in Phase~I, the time to a first answer starts peaking significantly above $7\,\unit{days}$ on average in Phase~II (largely q75-100 data points with fewer interactions below 1k interactions/week).
With the uptake in total system interactions, this value decreases for all community sizes.
Nonetheless, we overall observe huge variations in response times across the board.
Larger communities gradually maintain a 100-fold lower response time compared to their smaller counterparts.
This may be a primary driver for attracting more participation---the rich get richer, though larger communities widely also correlate to population figures.
We model this distribution with a shifted power-law at R2 scores of 0.23.
Note that our community-approximation may overestimate larger communities (if intercity distances are larger than the app radius); given this bias, content frequencies arguably still remain quite high.

\afblock{Conversationness.}
We measure conversationness as the number of discussion participants in a thread to the total amount of answers.
That is, values converging towards 0 represents only few participants, whereas values converging towards 1 denote a very wide audience participating within a thread.
We focusing an daily or weekly averages across all communities, we find huge variations in the conversationness.
On average, conversationness remains equal across community sizes at values of 0.66 with minor standard variation of up to 0.07 for both, daily and weekly aggregates.

To provide deeper insights into the structure of average conversations on the platform, in Figure~\ref{fig:conversationness} we show a scatter plot of conversationness to average thread length.
Again, we do not discover any difference between community sizes.
However, there is a natural trend to lower conversationness values with increasing thread length, \ie{} longer discussions likely are held between various participants.
Whereas shorter thread lengths up to five replies experience values above 0.65, we observe a rather linear decrease to only 0.55 for 15 replies.
Though a simple linear curve fits quite well, from our references, power-law (pwr) approximates this distribution best with an R2 score of 0.81 and better accounts for the heavy tail at shorter lengths.

\afblock{Community Vote Consensus.}
Finally, we want to provide deeper insights into the community voting behavior as is represents a vital factor for content appreciation and distributed moderation.
Previously in \sref{sec:interaction_dynamics}, we learned that overall community interactions are almost equally shared between creating content and voting.
Further, due to their exposure, especially threads are very likely to be upvoted; thus, cumulative scores are largely equal or above 0.
Yet, actual cumulative scores follow a power law being heavy tailed across posts (not shown).
That is many posts may only receive few votes---if any; only few will receive exceptional scores, promoted by tha app's \emph{loudest} feed. 

Given our observations of community voting behavior, we are further interested in homogeneity.
To what extent do users agree on dis- and liking contents, therefore do we find controversies in steering the communities?
As a measurement for vote consensus of a post, depending on which interactions are more dominant, we map the downvotes to overall votes to -1..0 and upvotes to overall votes to 0..1.
\Ie{} values converging towards -1 (1) denote dominating downvotes (upvotes); whereas both figures are equally cast around values of 0.
We provide a scatter plot of per city weekly consensus averages across communities in Figure~\ref{fig:vote_consensus}.
Note that we filtered out posts without votes.
The x-axis describes the amount of weekly community interactions, whereas the y-axis denotes the vote-censusus score as described; color represents community quantile.

As expected, due to the predominance of upvotes, consensus scores are widely on the positive side between 0.0 and 0.6.
\Ie{} about 70\% (90\%) upvotes may represent a consensus of 0.3 (0.6).
While we observe heavier skews in weekly averages for q25-50, there is an overall trend to increasing consensus scores in larger communities---rendering them more homogeneous.
For reference, the distribution fits a shifted power-law with R2 scores of 0.23.

\takeaway{
    \emph{i)} 
    Communities that have reached a certain critical user base show a lively behavior in terms of response times to posts--on average.
    Within larger communities, people experience a reply to their post within 10 to 20 minutes on average, while smaller communities reply orders of magnitude slower.
    A light correlation remarks a natural \emph{scaling effect} towards lower response times with increasing community activity.
    \emph{ii)} 
    \emph{Invariant} of community size, longer discussions naturally attract more participants.
    \emph{iii)} 
    This is also reflected in vote consensus scores on average largely being within the same positive regime---\emph{invariant} to community size.
    We find less variation in consensus values for larger communities that also tend to be more homogeneous.
}

\vspace{-1.5em}
\subsection{Modeling Community-User Activity}
\label{sec:model_user_activity}
To highlight the rapid growth of Jodel within the KSA, we next analyze and model driving and resulting factors over time proxied through the amount of registrations and interactions per community member.

\begin{figure}[t]
    \centering
    \begin{subfigure}[t]{.75\linewidth}
		\centering
        \includegraphics[width=\linewidth]{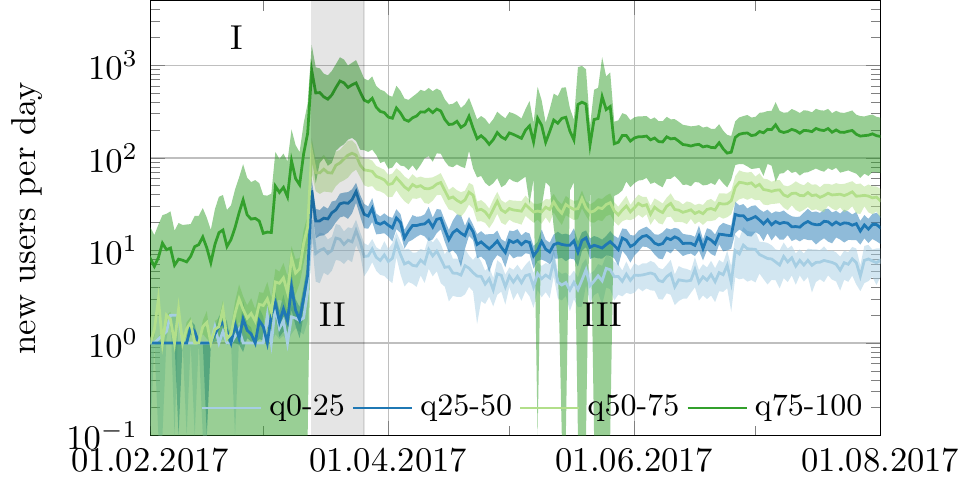}
        \caption{
            \textbf{Average \#registrations per day.}
        }
        \label{fig:spread_cities_registered__time}
    \end{subfigure}
    
	\begin{subfigure}[t]{1\linewidth}
		\centering
        \includegraphics[width=.75\linewidth]{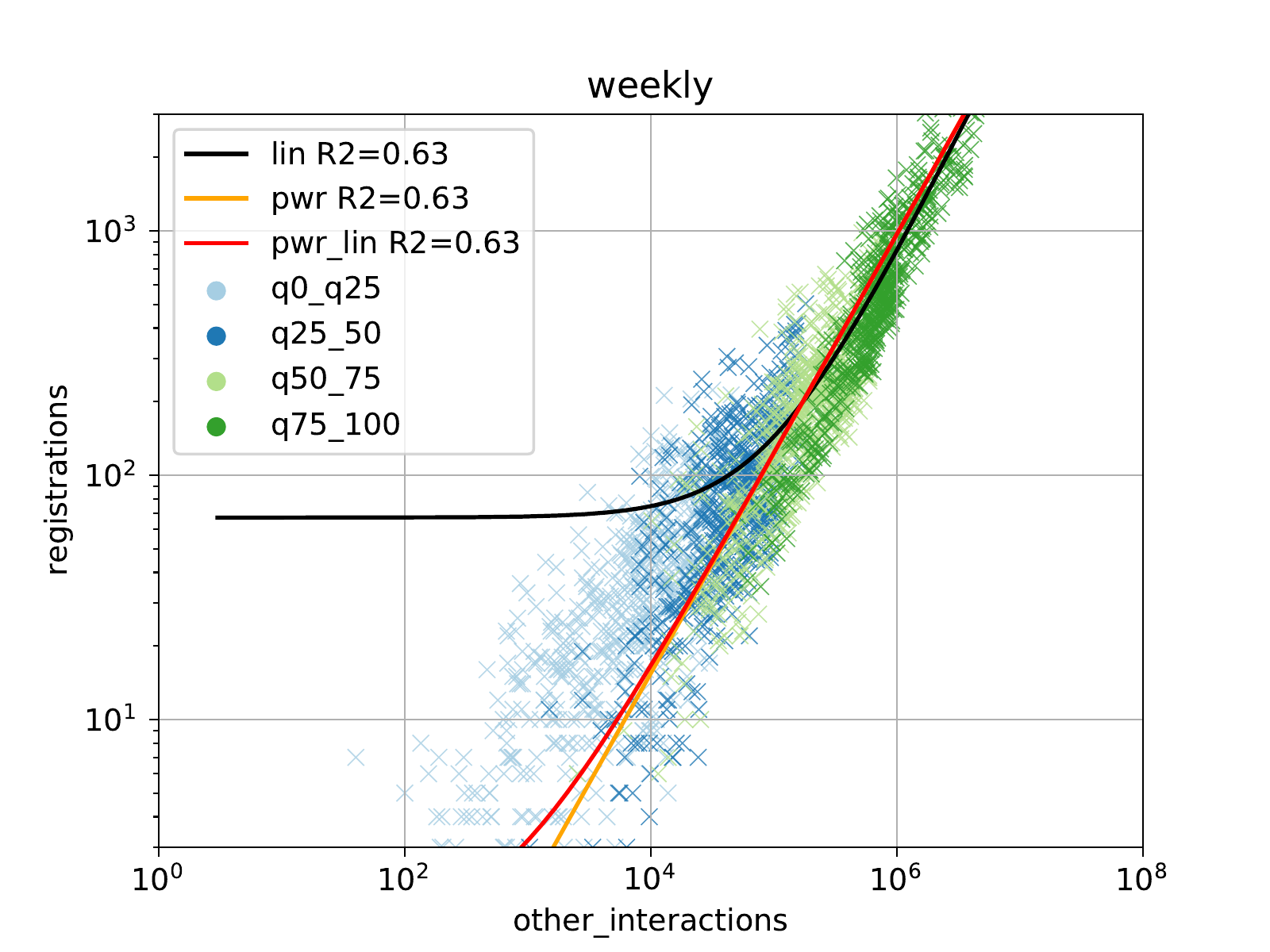}
        \caption{
            \textbf{Phase III model.}
        }
        \label{fig:spread_cities_registered__model}
    \end{subfigure}
    \caption{
        \textbf{Registration over time by city subgroup; modeling registrations to weekly city-interactions}.
        \emph{(a)} This figure describes average$\pm$std \#new user registrations per day per city quantile.
        \emph{(b)} This scatterplot denotes the correlation between new user registrations and other city interaction on weekly aggregates.
		A power-law fitting yields R2 scores of $0.63$, indicated by the linear distribution on the log-log plane.
    }
	\label{fig:spread_cities_registered}
    \vspace{-1.5em}
\end{figure}

\afblock{Registrations.}
In Figure~\ref{fig:spread_cities_registered__time}, we show the average number of new users per day (solid lines) and their standard deviation (shaded background) for groups of cities ranked by their number of system interactions (activity).
The figure shows four groups as percentiles according to community interaction volume. %
While the distribution varies heavily specially within Phase I \& II as indicated by the standard deviation, 
interestingly, all communities---irrespective of their size---show the same adoption pattern.
However, at steady state in Phase III, the registerations are \emph{scaled} by activity---an unexpected finding given the independence of the communities.
That is, the 3-phase adoption occurred at all communities simultaneously scaled by activity.
Here, the most active cities are characterized by the largest influx of new users, while the least active ones have the lowest influx.
The observation that all communities simultaneously followed the same adoption pattern supports our hypothesis that the adoption is triggered by external stimuli to simultaneously reach users nationwide.
Notably, the smaller communities experienced a larger influx of new users in phase II than larger communities indicating that more users were motivated to start using Jodel relative to their overall size.
Thus, the sudden adoption in March 2017 (Phase II) is characterized by a country-wide adoption of Jodel where each community experienced a substantial influx of new users.
Afterwards, in Phase III, each community experiences a rather constant influx of new users that only differs in absolute numbers relative to the size of the community.

To provide more evidence of this scaling effect in Phase III, we present cumulated registrations across weekly community interactions as a proxy for activity in Figure~\ref{fig:spread_cities_registered__model}.
This scatter plot's x-axis denotes weekly interactions per community, while the y-axis accounts for registrations.
Data point colors denote a city's quantile.
We observe a correlation between the number of newly registered users and the overall activity of each community.
That is, with increasing community activity, the communities also enjoy more and more new users onboarding.
The logarithmic representation may be deceiving in variability; in absolute terms, the standard deviation is almost as high as the mean registrations, regardless of aggregation time period (daily, weekly).
\Ie{} the std normalized by mean results in a preak ratio of 1.97 for the q75-100 quantile, wheres we find values of 0.76 to 0.97 for smaller communities on average.
Out of our fitted curves, all model the observed correlation at R2 scores of 0.63.
However, qualitatively, we would prefer power-law as it models the few interactions regime better.

\begin{figure}[t]
    \centering
	\begin{subfigure}[t]{1\linewidth}
		\centering
        \includegraphics[width=.75\linewidth]{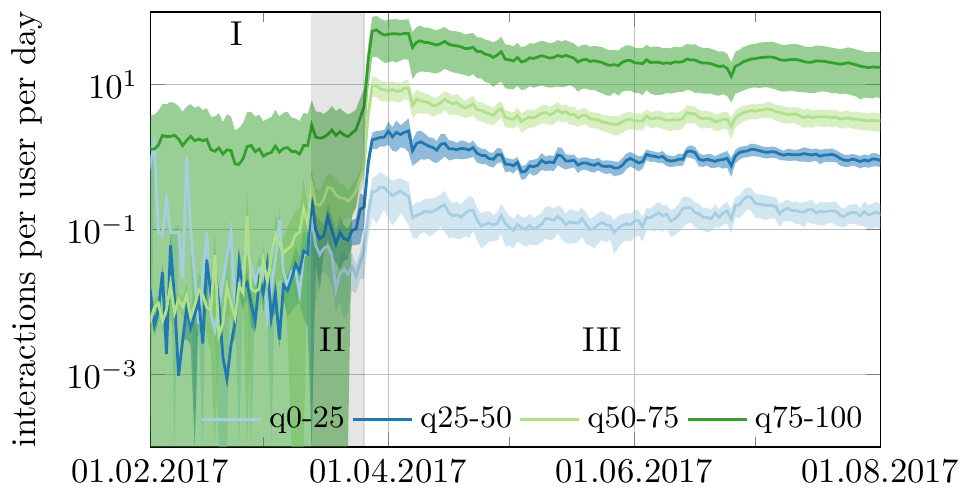}
        \caption{
            \textbf{Average \#interactions per user.}
        }
        \label{fig:spread_cities__time}
    \end{subfigure}

    \begin{subfigure}[t]{1\linewidth}
		\centering
        \includegraphics[width=.75\linewidth]{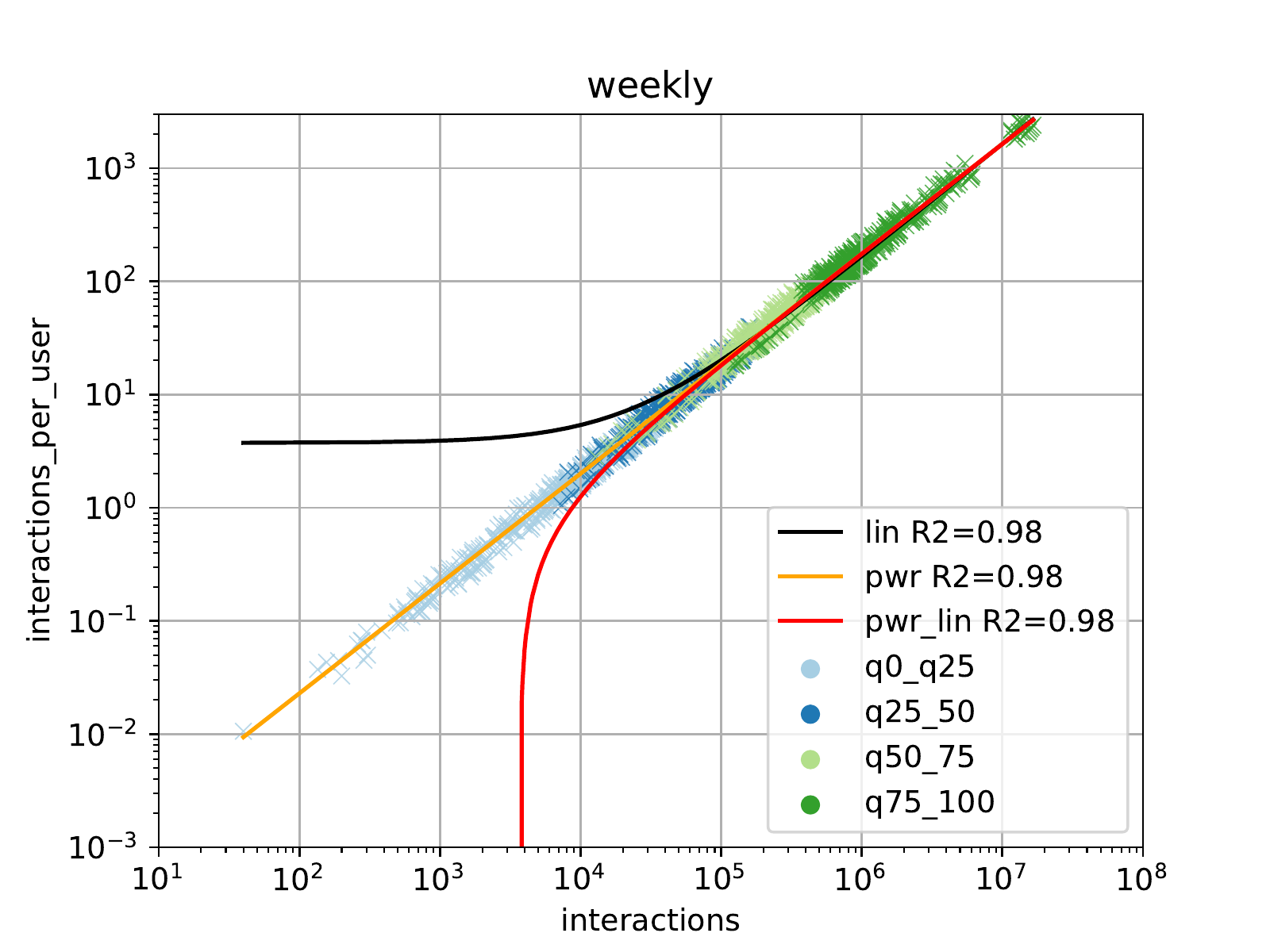}
        \caption{
            \textbf{Phase III model.}
        }
        \label{fig:spread_cities__model}
    \end{subfigure}
    \caption{
        \textbf{Interactions per user over time by city subgroup; modeling interactions per user to weekly city-interactions}.
        \emph{(a)} This figure describes average$\pm$std interactions per user per day per city quantile.
        \emph{(b)} This scatterplot denotes the correlation between per user interactions and total city interaction on weekly aggregates.
		A power-law fitting yields R2 scores of $0.98$, indicated by the strong linear distribution on the log-log plane.
    }
	\label{fig:spread_cities}
	\vspace{-1.5em}
\end{figure}

\afblock{Interactions.}
Before we focus the user perspective in the upcoming subsection, we already shift towards measures about users.
As we observe a steady influx of new users in especially within steady state Phase III, we wonder whether the amount of platform interactions per user follows the same pattern as the registrations---answering whether only more users of larger communities are responsible for more contributions, or if there are self-reinforcing effects at play as well---the rich get richer.

That is, we start with the amount of community interactions over time in Figure~\ref{fig:spread_cities__time}.
The x-axis show time, whereas the y-axis denotes the number of interactions per user.
We plot the average amount (solid lines) and the corresponding standard deviation (shaded background) for all community quantiles.
As observed for registrations, we largely find a qualitatively equal picture.
As seen before (\sref{sec:temporaladoption}), except for few larger communities, there is only little activity in Phase I \& II.
Similar to registrations, the volume of interactions over time remains stable and the quantiles equally tear apart with reasonable margins.

For better understanding this scaling effect between community size and per user interactions, we next provide a scatter plot showing community averages over time across both dimensions in Figure~\ref{fig:spread_cities__model}.
The x-axis represents the amount of weekly interactions per community; the color denotes the community quantile.
On the y-axis, we show the amount of interactions per user.
Note that we discuss the (power-law) per user interaction distributions in more detail later (\sref{sec:user_interactions}).
We again observe a clear correlation between both dimensions with only little variation, which is very similar to our findings in registrations.
This distribution can be modeled eceptionally well with all our fitting approaches with R2 scores of 0.98; however, the lower regime is qualitatively represented best with a  shifted power-law.

\takeaway{
    \emph{i+ii)} Geographically, \emph{invariant} to communities size, observed registrations and interactions follow the same qualitative behavior.
    Surprisingly, the massive influx of new users in March (phase II) occurred in simultaneously nationwide; all communities first peak in registrations and interaction peaks follow two weeks later. 
    This supports the earlier hypothesis that the adoption in phase II is likely triggered externally.
    Amounts of registrations and per user activity follows a power-law over the total community activity, a \emph{scaling effect}.
}

\vspace{-2em}
\section{A User-centric View}
\label{sec:user}

After having investigated the adoption pattern and provided an empirical overview of interlinked community interactions.
However, up to this point, we are missing the important complementary user perspective.
Thus, we next set out to characterize user behavior \wrt{} different communities in detail, and discuss metrics capturing the app's key design features.

\vspace{-1em}
\subsection{User Interactions}
\label{sec:user_interactions}

\begin{figure}[t]
    \centering
    \begin{subfigure}[t]{.475\linewidth}
		\centering
        \includegraphics[width=1\linewidth]{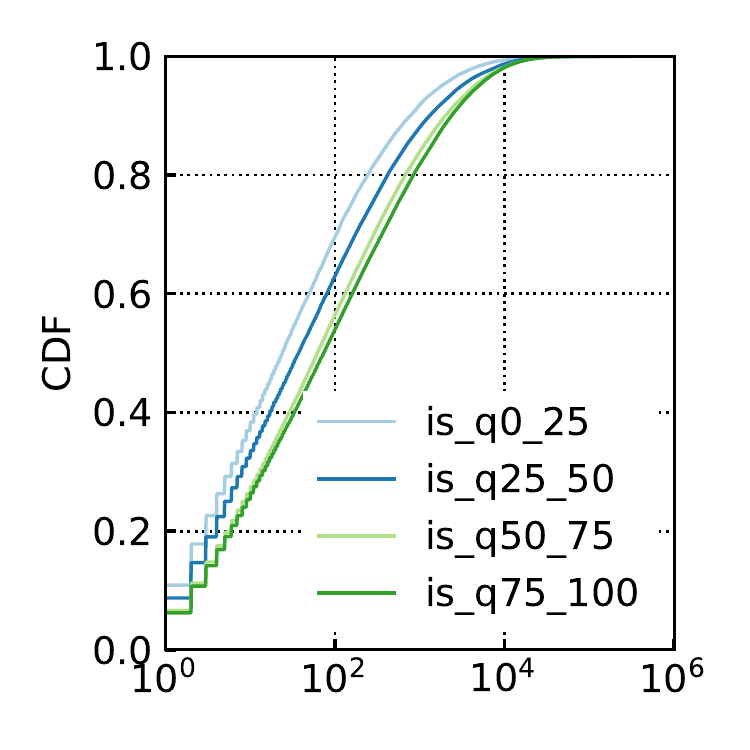}
        \caption{
            \textbf{User Interactions.}
        }
        \label{fig:cdf_user_interactions}
    \end{subfigure}
	\begin{subfigure}[t]{.475\linewidth}
		\centering
		\includegraphics[width=1\linewidth]{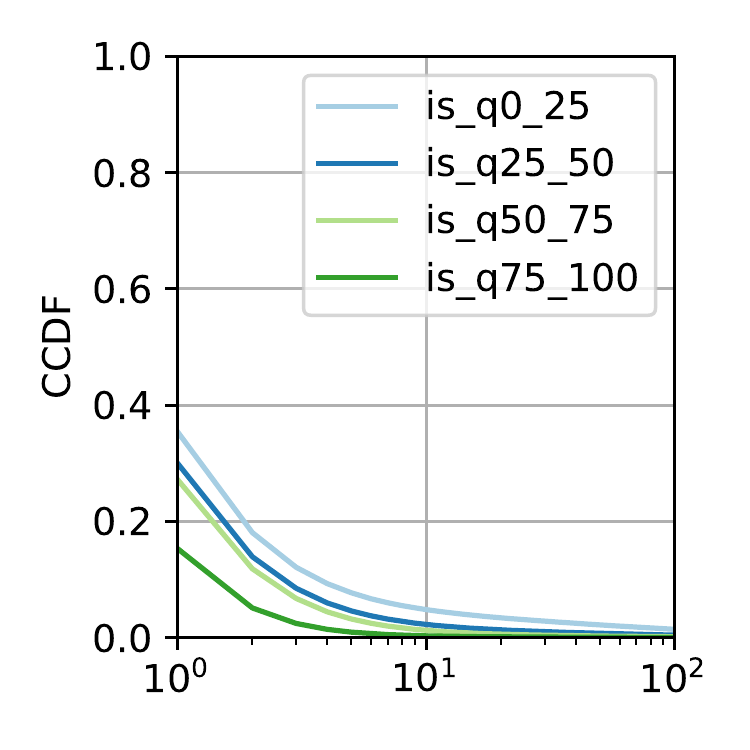}
        \caption{
            \textbf{User Re-Encounters.}
        }
        \label{fig:cdf_encounters}
    \end{subfigure}
	\vspace{1em}
    \caption{
        \textbf{User Interactions \& Anonymity in a nutshell.}
		\emph{(a)}~Interactions per user are power-law distributed.
		Most users only rarely participate, fewer others engage the platform more often.
		\emph{(b)}~Most user encounters across threads are unique; thus re-seeing another user is unlikely with exponentially decreasing probability, multi-encounters become less likely in larger communities.
    }
	\label{fig:cdf_user_interaction_encounters}
	\vspace{-1.5em}
\end{figure}

First, we want to clear the stage by showing per user platform interactions as a cumulative distribution function (CDF) in Figure~\ref{fig:cdf_user_interactions}.
While the logarithmic x-axis denotes the amount of total user interactions, the y-axis represents the accumulative fraction of users.
We plotted the CDF for each community quantile.
About 10\% of all users only opened the app (registration event) and never actively participated.
We further find that depending on community size, 50\% within larger to 70\% within smaller communities, the users only interacted up to 100 times with the platform.
The CDF indicates a power-law distribution, which a linear shifted power-law curve fits very well with R2 scores of 0.98---however, the heavy tail experiences a drop-off not being modeled well (now shown).

\takeaway{
	\emph{Invariant} to community size, we find the wide majority of users browsing rather casually.
	Only few power users contribute overbalanced.
}

\vspace{-1em}
\subsection{Anonymity - Absent Social Ties}

Social Networks have been shown over and over again to form small-world connection graphs, \ie{} a user probably knows at least some friends of friends while keeping an overall small graph diameter.
Such social graph depend on social structure and ultimately some kind of user profile reflecting social credit.
This however is not possible in a completely anonymous environment like Jodel.

To show that communication on Jodel is very ephemeral, we determined how often users encounter each other by replying on one's thread.
We show our results as a complementary cumulative distribution function (CCDF) across encounters distinguished by community quantile in Figure~\ref{fig:cdf_encounters}.
The logarithmic x-axis denotes encounters, \ie{} how often users re-interact with each other within any other thread.
On average, about 83\% (10\%) of all encounters remain unique (re-encountered once*).
For more encounters, we observe an exponential decrease in occurrences, such that two (three) encounters happen across in 10\% (3\%) of all encounters.
However, due to their volume, q75-100 values introduces a skew towards higher values dat 84\% (10\%) unique* encounters.
We observe a shift of fewer encounters with increasing community activity.
That is, in smaller q0-25 communities, only 62\% (17\%) encounters remain unique*.

\takeaway{
	We find a natural \emph{scaling effect} in increasing community activity leasing to users being less likely to interact with the same person ever again---rendering platform communication rather ephemeral.
}

\vspace{-1em}
\subsection{Hyperlocality - User Communities}
\label{sec:user_locations}

\begin{figure}[t]
    \centering
	\begin{subfigure}[t]{.475\linewidth}
		\centering
		\includegraphics[width=1\linewidth]{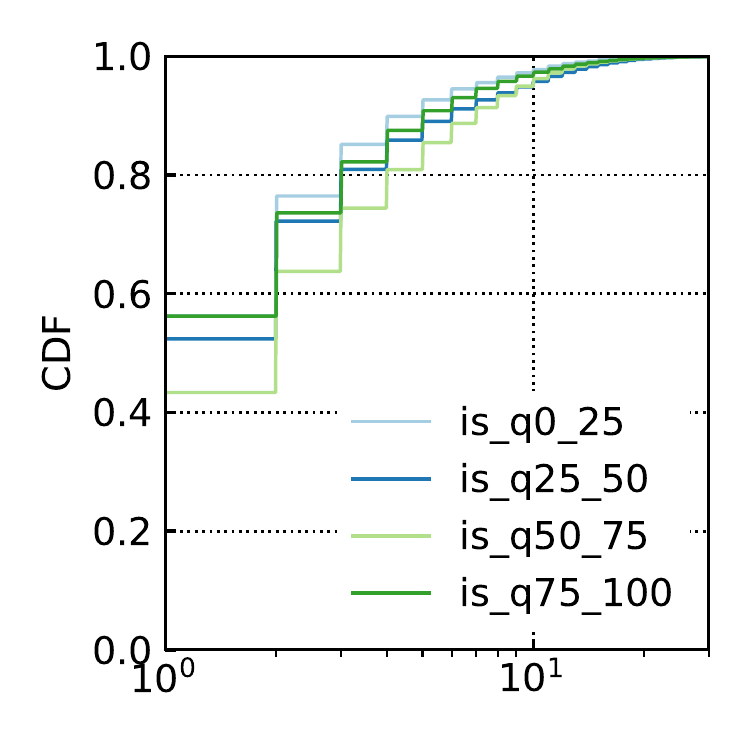}
        \caption{
            \textbf{Number of Communities per User.}
        }
        \label{fig:cdf_location_num}
    \end{subfigure}
    \begin{subfigure}[t]{.475\linewidth}
		\centering
        \includegraphics[width=1\linewidth]{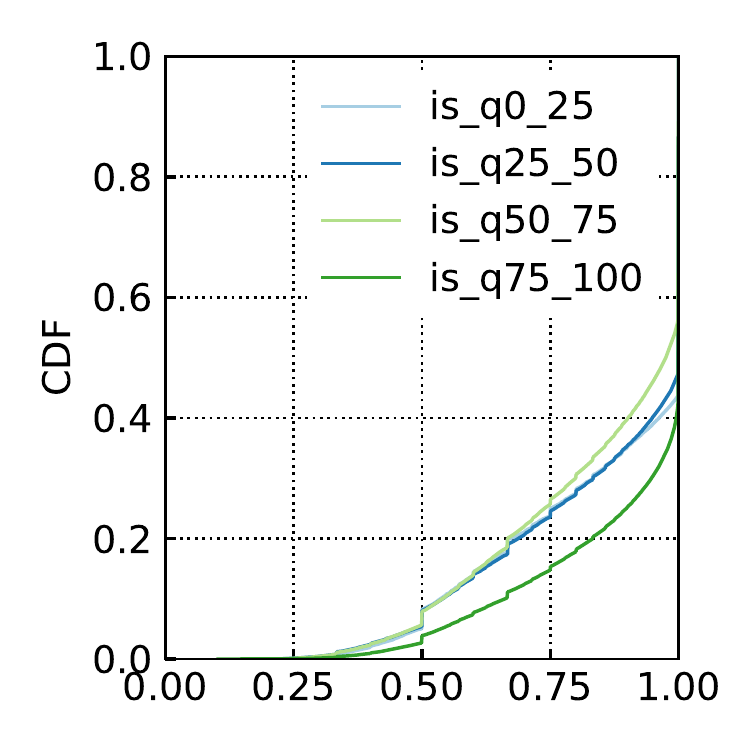}
        \caption{
            \textbf{Top1 Community Interaction fraction.}
        }
        \label{fig:cdf_location_ratio}
    \end{subfigure}
    \caption{
        \textbf{Hyperlocality in a nutshell.}
		\emph{(a)} Most users only ever participate in a single community with a steep decrease.
		Observations remain similar across community sizes.
		\emph{(b)} Users having participated in multiple communities still focus on a single home community. %
		Users of larger communities tend to share more interactions on their home community.
    }
	\label{fig:cdf_app_features_nutshell}
	\vspace{-1.5em}
\end{figure}

While we have previously elaborated on anonymity as a central app design feature, we now want to shed light on the other property of hyperlocality.
That is, we explore the amount of communities a user interacts with, and further evaluate to which extent users focus their content on their favorite community.

We show cumulative distrbution functions (CDFs) for the amount of communities a user has ever interacted with in Figure~\ref{fig:cdf_location_num}, distinguished by the user's home community quantile.
While we do no observe considerable differences across community sizes, most users stick to a single community.
The amount of communities per user rapidly decreases, such that about 75\% of all users participate in up to two communities.

To provide a deeper insight to which extent users distribute their activity across communities, we additionally measure the fraction of a user's home community (Top1) as a CDF in Figure~\ref{fig:cdf_location_ratio}.
As seen before, depending on community size, 40-60\% of all users users participate only in a single communitiy.
However, users having participated in more than one community still focus on their home community, \ie{} about 20\% of users distribute less than
25\% across others. 
Further, we observe that users in larger communities tend to share more interactions on their home community; this might be a result of reported hints to use fake-GPS for joining Jodel in larger communities being believed to be \emph{better}.

\takeaway{
	Most users participate only in a single community.
	Those with multiple communities still focus on their home-community.
}

\subsection{Modeling Active Users \& Lifetime}
\begin{figure}[t]
	\begin{subfigure}[t]{1\linewidth}
		\centering
        \includegraphics[width=.75\linewidth]{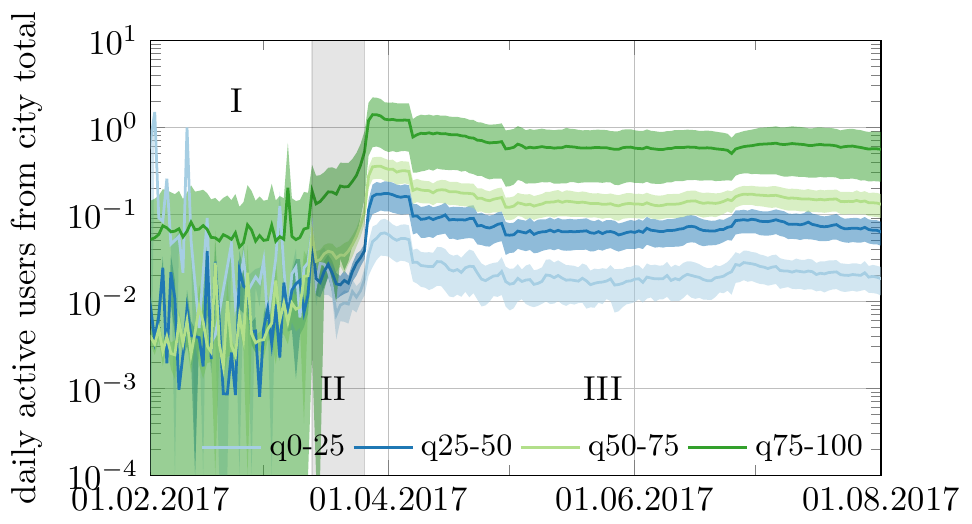}
        \caption{
			\textbf{Ratio of daily active users.}
		} 
        \label{fig:t_daily_active_users__time}
    \end{subfigure}

	\begin{subfigure}[t]{1\linewidth}
		\centering
        \includegraphics[width=.75\linewidth]{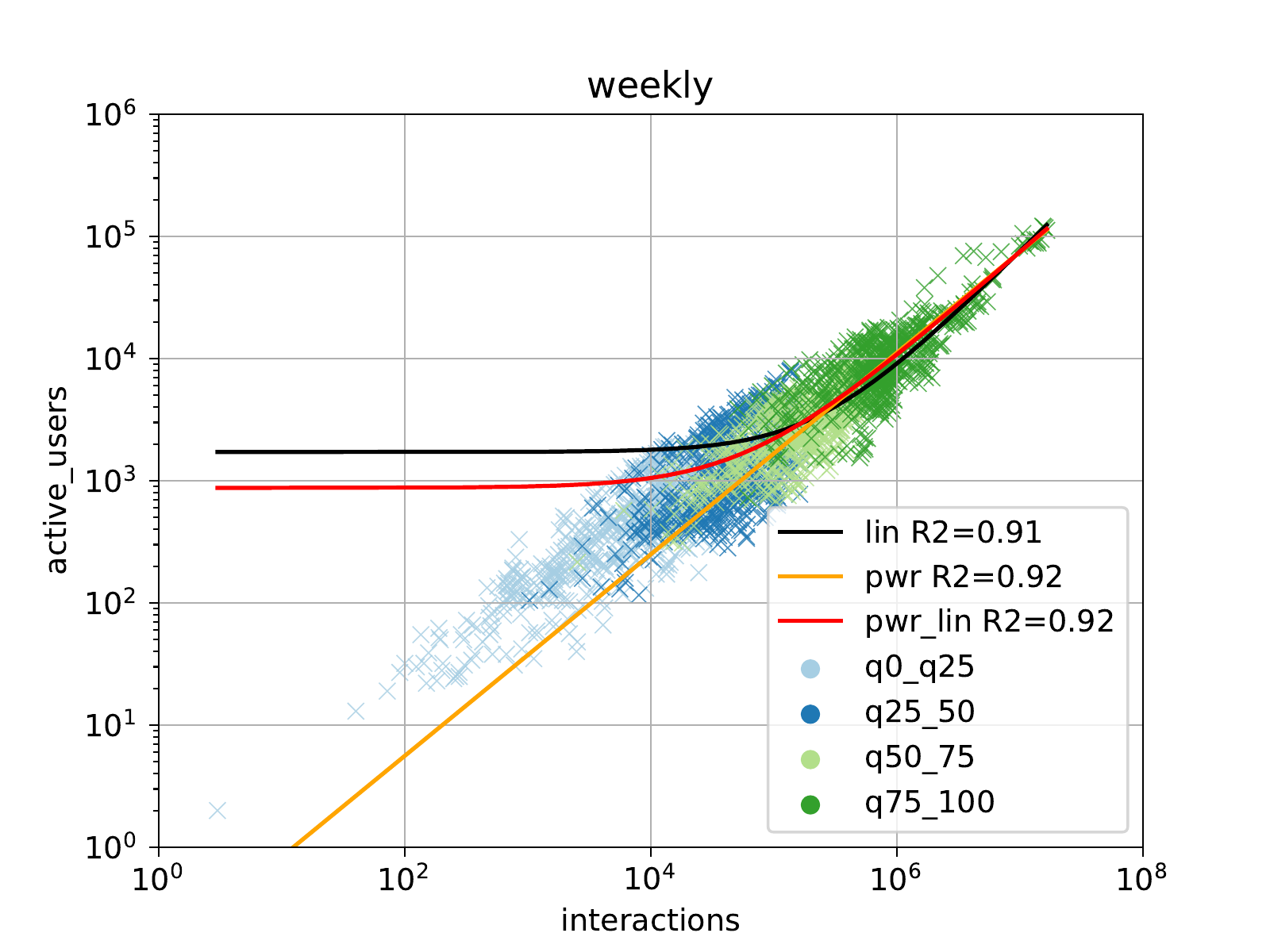}
        \caption{
            \textbf{Phase III model.}
        }
        \label{fig:t_daily_active_users__model}
    \end{subfigure}
    \caption{
        \textbf{Daily active users over time by city subgroup; modeling daily active users to weekly city-interactions}.
        \emph{(a)} This figure describes average$\pm$std active users per day per city quantile.
        \emph{(b)} This scatterplot denotes the correlation between active users and total city interaction on weekly aggregates.
		A (linearly shifted) power-law fitting yields R2 scores of $0.92$, indicated by the strong linear distribution on the log-log plane.
	}
	\label{fig:t_daily_active_users}
	\vspace{-1.5em}
\end{figure}

Previously, we provided community driven user-aggregate measures in interactions and onboarding.
In this section, we will complete this picture discussing daily active users of communities and user lifetime.

\afblock{Daily Active Users.} 
We first study the active user-base of each community---the daily active users---and if communities differ by those.
$14.6\%$ of all registered users never interacted with the Jodel app and 27.6\% drop out within 24 hours.
Since pure registration figures (\sref{sec:model_user_activity}) do not represent the active users, we next focus on users that interact with Jodel, \ie{} the active user base.
We show the amount of daily active users in Figure~\ref{fig:t_daily_active_users__time}, again aggregated into the four community subsets normed by the amount of registered users by each point in time. %
While the x-axis denotes the temporal axis, the log-scaled y-axis reveals the ratio of users having interacted with the system on a particular day at least once to total city registrations up to this point in time.
The shaded areas denote the standard deviation within each city subset.

While there is primarily noise in Phase I, given the low number of users, the daily active user base grows significantly across all communities after the initial heavy user influx in March (Phase II).
At the beginning of April (Phase III), this value decreases slightly, but stabilizes afterwards and remains on the same level until the end of our observation.
We clearly observe the same qualitatively development across all community sets, varying in absolute numbers by orders of magnitude.
\Ie{} users more likely increase their attention to larger communities.

We further provide a scatter plot of weekly aggregates of active users across communities in Figure~\ref{fig:t_daily_active_users__model}.
While the x-axis represents the weekly interactions, the y-axis accounts for the amount of active users.
Our curve fittings indicate that the distribution follows our curve fitting approaches almost equally well with R2 scores of 0.92, a shifted power law qualitatively describes observed behavior best.

\afblock{User Lifetime.} 
The previous analysis showed that larger communities attract more users to participate reflected in the daily active users. %
Yet, we do not know how this translates into the time users stick to the system, which is why we now evaluate the time for how long users keep using the app.
There is a variety of possibilities in the extremes of \emph{a)} a cyclic renewal of the complete user base happens over and over again, or \emph{$z$)} users are very committed to their community and participate over longer time periods.

\begin{figure}[t]
		\centering
		\includegraphics[width=\linewidth]{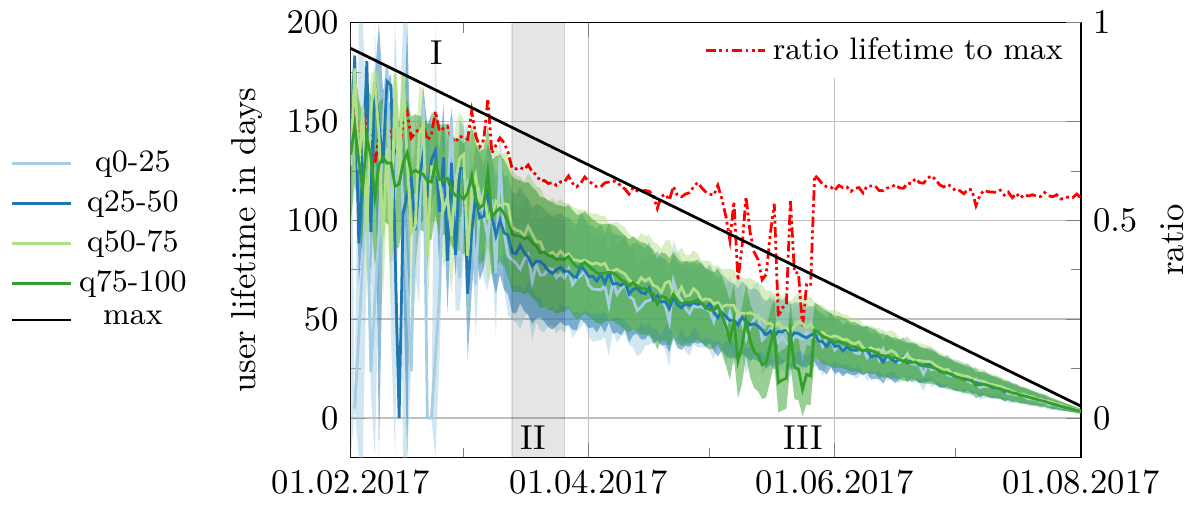}
		\caption{
			\textbf{Users' lifetime.} 
            This figure shows the average user lifetime and stddev by their registration date.
            While being noisy in Phase I, this metric stabilizes equally for all city subsets later, depicted by the lifetime ratio to the upper bound (end of observation) at a constant level above 55\%.
		}
		\label{fig:t_active_duration}
        \vspace*{-1.5em}
    \end{figure}

To answer this question, we show the user lifetime within the system in Figure \ref{fig:t_active_duration}.
This figure shows the average lifetime and standard deviation of users w.r.t their registration date.
While the x-axis denotes time, the y-axis marks the number of days a user is active (registration to last system interaction).
Due to the end of our observation period, the active days are bounded (max).
The ratio lifetime to max resembles the fraction of overall average user lifetime to this bound.

We make three observations:
\emph{i)} We observe that the user lifetime is quite high, but noisy in Phase~I, while decreasing within the inception Phase~II.
Then, it stabilizes within Phase~III indicated by the linear trend of the user lifetime.
\emph{ii)} The ratio of the overall user lifetime to the given observation bound indicates that on average more than $60\%$ of the users keep using the app until the end of observation.
\emph{iii)} There is no qualitative difference between community sizes as although, there are huge differences in absolute numbers, the measured user lifetime is rather identical---a similarity.

\takeaway{
	\emph{i)} 
	With increasing community activity, the amount of daily active users also increase following a power-law. 
	The amount of active users is in a steady state for Phase III as the communities do not differ qualitatively.
	\emph{ii)} 
	There is high user retention indicated by $>60\%$ of the users keeping using the app until the end of our observation.
	Though user lifetime fluctuates, on average, it remains qualitatively similar throughout the city quantiles.
	We conclude that many users stick to the system, while there also happens a cyclic renewal of the user base for the remaining $40\%$ users.
}

\vspace{-1em}
\subsection{User Retention \& Churn}
\label{sec:churn}
Next, we study what makes users to continue using Jodel (retention) or to leave the app (churn).
That is, we present a similarity across all communities to study if can we characterize long-time users from their first two days in the platform?
Further, we take a closer look into differences of churned users between community sizes from a network perspective.

While the population of user interactions that have dropped out in various time frames between communities has no statistical significant difference, we find
that 60\% of the users keep using the app after registration, 
However, about $15\%$ of dropped users
do not interact with the platform at all.
From our data, we cannot assess whether these users simply do not use the app, or whether they are lurkers who entirely consume content entirely \emph{passive}.

\begin{figure}[t]
	\centering
	\includegraphics[width=\linewidth]{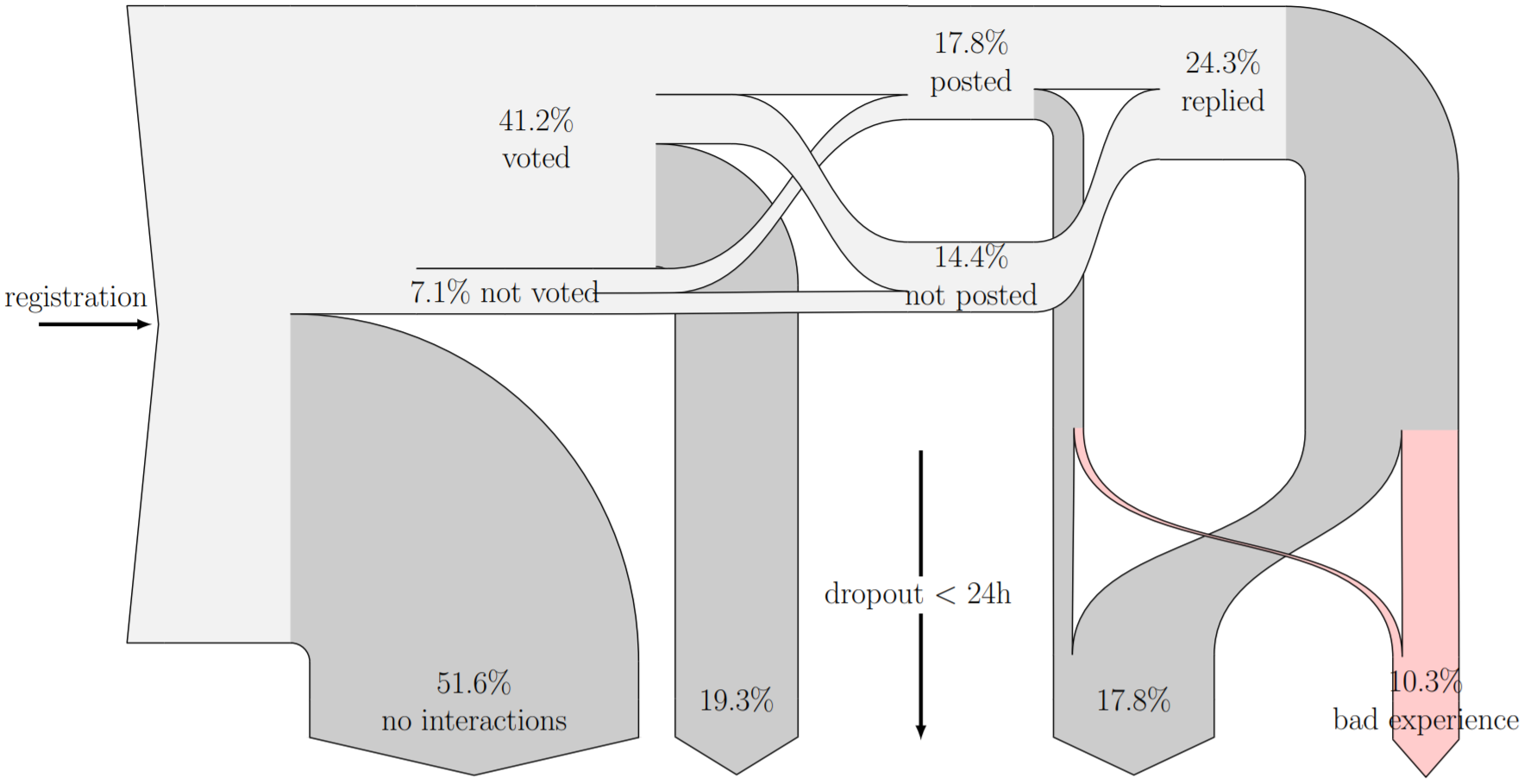}
	\caption{
		\textbf{Qualitative view of overall lost users within 24h.}
		This Sankey diagram shows actions of all users before dropping out of the system.
		Most users do not interact at all (no interaction), while the others at least vote once (voted) before dropping out.
		A \emph{bad experience} means getting downvotes on posts.
	}
	\label{fig:dropped_sankey}
\end{figure}

\begin{table}[t]
	\small
	\centering
	\resizebox{\linewidth}{!}{
		\begin{tabular}{|l|r|r|r|r|r|}
			\hline
			\textbf{comm.}	& \textbf{no interact.}	& \textbf{voted}		& \textbf{posted}	& \textbf{replied}\\ \hline\hline 		%
			q0-25					& 50.1$\pm$5.6\%			& 38.5$\pm$7.0\%	& 18.7$\pm$3.6\%	& 22.5$\pm$3.7\%\\ \hline	%
			q25-50					& 43.7$\pm$4.7\%			& 46.2$\pm$5.4\%	& 19.8$\pm$3.3\%	& 27.2$\pm$4.1\%\\ \hline	%
			q50-75					& 42.1$\pm$5.0\%			& 48.9$\pm$5.3\%	& 20.3$\pm$2.7\%	& 28.1$\pm$4.3\%\\ \hline	%
			q75-100					& 41.8$\pm$11.0\%			& 50.1$\pm$9.6\%	& 21.2$\pm$4.1\%	& 29.5$\pm$5.8\%\\ \hline	%
		\end{tabular}
	}
	\caption{\textbf{Qualitative view of users within 24h by community sizes.}
		This table shows the average amount (and stddev) of performed interactions of users that dropped out within 24h.
		There is a clear tendency of users participating more in larger communities.}
	\label{tab:user_churn}
	\vspace{-2em}
\end{table}

\afblock{User Retention.}
To study what differentiates users that kept using the app (retention) from users that dropped out (churn), we focus on their interactions with the app.
We take users registered after April 1 (Phase III, nationwide establishment) and group them into three groups: \emph{i)} users that were active only for two days, \emph{ii)} only for a week, and \emph{iii)} users that kept using Jodel for more than 30 days.
For each group, we extracted the amount of interactions of each user on the Jodel platform within the first 24 hours after registration and determined the user's community set.

First, we compare the total number of system interactions (\ie{} creating content and voting) between the different groups and communities.
We observe that the user populations do {\em not} significantly differ (tested with a t-test).

Second, we analyze differences \wrt{} different interaction types (\ie{} posting, replying, up- and downvoting, and flagging) in isolation, we arrive at the very same result (again with a t-test): there is no significant difference between our defined groups---a similarity across all community sets.

\afblock{User Churn.}
    There is no obvious difference in the populations of user interactions \wrt{} retention.
    We next flip the question and study why and how users churn (leaving the app).
    We begin by studying users that dropped out of the system, \eg{} by losing interest.%
	To shed light on this group of users, we analyze behavioral metrics of users who did not interact with the system at all or their lifetime was limited to only at most 24 hours.

	First, in Figure~\ref{fig:dropped_sankey}, we provide a high-level view on the dropped user base via a Sankey diagram (describing qualitative flows).
	Most dropped users (about 51.6\%, 174k) have no interaction with the system at all.
	These users installed the application and opened it at least once to trigger a system registration, but did not actively interact by posting, replying, or voting.
	From our data, we cannot tell whether these users did not use the application at all or used it only by means of passively consuming content (\ie{} browsing over and reading posts).
	Most other dropped users at least voted once, while replying is more popular than posting among them.
	Still, about 19.3\% of the dropped users only voted, while the others created content.
	Out of these content creators, we counted the users having a \emph{bad experience}, that is getting downvotes to one of their posts or even getting blocked by moderation---accounting for 10.3\% of all dropped users.

	Second, we split the users into community subsets and analyze the non-/presence of possible interaction types.
	We provide the results in Table~\ref{tab:user_churn}.
	It provides information of the average amount (and standard deviation) of how many users of each community subset have either not interacted with the platform at all, voted, posted or replied.
	For all community subsets, we observe a difference in these figures.
	That is, in smaller communities, more people have not interacted at all and other interaction types are less common than in larger communities.
	On the contrary, larger communities (in terms of interaction volume) trigger more users to interact.
	As we observe positive trends on average across community sizes.
	However, given overlapping standard deviations across averages arguably represents rather similar behavior. %

\takeaway{
	\emph{i)} 
	All communities show similar behavior \wrt{} user retention, an \emph{invariant}.
	That is, all communities behave similar by their interaction volume and interaction types subject to users lifetime.
	\emph{ii)}
	27.6\% of all registered users drop out within 24 hours.
    Although about 50\% of the users interacted with their community by voting or posting, half of them created content at least once (25\%) of which only about 10\% actually make a bad experience from an empirical point of view.
    \emph{Invariant} to community size, churned users behave similar before leaving the platform. %
}

\vspace{-1em}
\section{Related Work}
	\label{sec:rw}
	Social network analysis is an active field of research for more than a decade.
	Research provided a general understanding through the empirical and qualitative analyses of a number of different networks.
	Examples include structural measurements classic online social networks (\cite{mislove2007measurement, nazir2008unveiling, schioberg2012tracing,kairam2012talking}) as well as more specialized variants such as microblogging (\cite{bollen2011modeling}), picture sharing (\cite{vaterlaus2016snapchat, cha2009measurement}), or knowledge sharing (\cite{wang2013empirical}).
	Works in this field analyzed the networks' {\em structure}, mostly using graph-theory approaches. %
	This way, they have shown that social networks usually create small-world networks (\cite{manku2004know,freeman2004development}).
	Besides structural network properties, qualitative analyses focused on content.
	Examples include sentiment analysis (\cite{thelwall2010emotion,kouloumpis2011twitter,tan2011user,bermingham2009combining}).
	Further content analysis targeted the (geospatial) information spreading (\cite{fink2016investigating,kamath2013spatio,yin2011geographical}), or Youtube (\cite{brodersen2012youtube}).

	Little is known about the early adoption of a new social network.
	Existing works partially provide information about the growth and development of online social networks, such as Yahoo 360 and Flickr (\cite{kumar2010structure,mislove2008growth}), Google+ (\cite{gong2012evolution,schioberg2012tracing}), Facebook (\cite{wilson2009user}) and others (\cite{benevenuto2009characterizing,jiang2013understanding}) usually relying on sampled information; further, they do not focus on drivers or reasons for network growth in particular.
	Especially our ground truth information enables us to empirically trace the birth and development of a new community in detail---in this case of the KSA, an external trigger, \ie{} other social media, has turned Jodel on in a sigmoid fashion.

	Another line of research studied anonymous usage.
	Anonymous usage can not only catalyze adverse content (\eg Yahoo Answers \cite{kayes2015social}), but also cyber-bullying (\cite{whittaker2015cyberbullying}) \eg \mbox{Ask.fm} (\cite{hosseinmardi2014towards}).
	Countermeasures usually rely on filtering, \eg community-driven as for Reddit (\cite{van2011human}).
	More broadly, steering of user behavior is a topic on its own for, \eg Q\&A platforms (\cite{anderson2013steering,grant2013encouraging}).
	Although anonymous networks can struggle with negative content, they have shown to serve a demand, \eg semi-anonymous confession boards (\cite{birnholtz2015weird}) or anonymous platforms (\cite{stutzman2013silent,bernstein20114chan}).
	Some users even tend to create throwaway accounts (\cite{leavitt2015throwaway}).
	Apart from online social networks, there are several applications that enable anonymous posting.
	Whisper enables users to interact anonymously on a global basis.
	It was empirically studied by \cite{wang2014whispers}, with a distinct focus on classifying the anonymity sensitivity of the posted content (see \cite{correa2015many}).
	The location-based application YikYak (similar to Jodel) was empirically studied by~\cite{mckenzie2015oxen}, with a focus on classifying the exchanged content, see~\cite{lee2017people,black2016anonymous}.
	We complement these works by contributing the first large-scale empirical analysis of the Jodel messaging application in a unique view based on complete ground truth information provided by the network operator itself.

\vspace{-1em}
\section{Conclusion}
\label{sec:conclusion}

In this paper, we empirically characterize the sudden nation-scale adoption of Jodel---a location-based and anonymous messaging application---in the KSA: from the first post to saturation through the lens of an operator by using complete and ground truth data.
The location-based nature of Jodel forming hundreds of {\em independent} communities throughout a complete country enables us to compare their adoption patterns.
The major adoption phase is characterized by a sigmoid nation-wide user growth and a two week delayed startup in user activity in \emph{all} communities.
That is, the adoption is characterized by a massive influx of users that occurred in all communities nationwide.
We hypothesize that this adoption must have been triggered externally by other social media, such as Twitter or Instagram, and is not the result of organic growth or epidemic spread.

By comparing these communities \wrt{} interaction volume (size), we identify similarities, (power-law) scaling effects in community size and rare differences.
However, we identify scaling effects: larger communities attract more users to be active on a daily basis.
Also, independent of community size, the observed amount of un- and popular content as well as the ratio of upvotes (happyratio) is similar across all city sizes.
Social credit is granted within minutes in larger communities (reply to a post) while being orders of magnitude slower in smaller cities, scaling with size.
We further identify that content voting popularity differs between the city subsets: users in larger communities are more likely to start new threads in comparison to smaller communities, although there already is a substantial amount of content available to them.
While we find similarities between the community sizes in user lifetime and retention, regardless of community size, positive reactions correlate with a user's lifetime and her number of interactions.
Yet, invariant to their size, all communities develop a stable daily active user base with more than 60\% of of the users keeping using the app until the end of our observation on average.

In future work, it would be interesting to complement this empirical work with a content perspective---not only within the KSA.
A further relevant study is to derive if these adoption effects are controllable, \ie{} can they be applied to a new country?
However, we consider further investigation of platform design decision most important---how does anonymity paired with location-basedness impact user experience?

\balance

\end{document}